\documentclass[12pt]{article}\usepackage{amsmath,amsfonts, epsfig, setspace, lscape}
 \begin{document}
 \def\const{\hbox{\rm const}}
 \def\diag{\hbox{\rm diag}}
 \def\mod{\hbox{\rm mod}}
 \def\rank{\mathop{\hbox{\rm rank}}}
 \def\Image{\mathop{\hbox{\rm Image}}}
 \def\mes{\mathop{\hbox{\rm mes}}}
 \def\grad{\mathop{\hbox{\rm grad}}}
 \def\ind{\mathop{\hbox{\rm ind}}}
 \def\exp{\mathop{\hbox{\rm exp}}}
 \def\d{\mathop{\hbox{\rm d}}}
 \def\k {\mathop{\hbox{\rm k}}}
\def\Diff{\mathop{\hbox{\rm Diff}}}
 \def\ad{\hbox{ad}}
 \def\Ker{\mathop{\hbox{\rm Ker}}}
 \def\Im{\mathop{\hbox{\rm Im}}}
 \def\id{\mathop{\hbox{\rm id}}}
 \def\Tr{\mathop{\hbox{\rm Tr}}}
 \def\so{\mathop{\hbox{\rm so}}}
 \def\Re{\mathop{\hbox{\rm Re}}}
 \def\Im{\mathop{\hbox{\rm Im}}}
 \def\scirc{\mathbin{\hbox{\scriptsize$\circ$}}}
 \def\M{\mathop{${\mathbb R}^2_1$}}
 \def\sgn{\mathop{\hbox{\rm sgn}}}
 \def\arctanh{\mathop{\hbox{\rm arctanh}}}
\def\bX{\mathop{\hbox{\bf X}}}
\def\bV{\mathop{\hbox{\bf V}}}
\def\bK{\mathop{\hbox{\bf K}}}

\newtheorem{con}{Conjecture}[section]
\newtheorem{prob}{Problem}[section]
 \newtheorem{prop}{Proposition}[section]
 \newtheorem{coro}{Corollary}[section]
 \newtheorem{lem}{Lemma}[section]
 \newtheorem{theo}{Theorem}[section]
 \newtheorem{defi}{Definition}[section]
 \renewcommand{\theequation}{\thesection.\arabic{equation}}
 \newtheorem{cri}{Sufficient condition}
 \renewcommand{\thecri}{\arabic{cri}.}
 \newenvironment{criterion}{\begin{cri}\rm}{\end{cri}}
 \newtheorem{rem}{Remark}[section]
 \newtheorem{exa}{Example}[section]
 \newtheorem{hyp}{Hypothesis}[section]
 \newenvironment{hypothesis}{\begin{hyp}\rm}{\end{hyp}}
 \renewcommand{\therem}{\thesection.\arabic{rem}}
 \newenvironment{remark}{\begin{rem}\rm}{\end{rem}}
 \newfont{\gothic}{eufm10 at 12pt}
 \title{Covariants, joint invariants and the problem of equivalence  in the invariant 
 theory of Killing tensors defined in pseudo-Riemannian spaces of constant curvature}
 \author{Roman G. Smirnov\footnote{Electronic mail: smirnov@mathstat.dal.ca} and Jin Yue\footnote{Electronic mail: 
 jyue@dal.ca} \\
   Department of Mathematics and Statistics, 
Dalhousie University \\ Halifax, Nova Scotia,  Canada B3H 3J5 }
    \date{ }
 \maketitle
\noindent Running title: {\large Covariants and joint invariants of Killing tensors}

\doublespacing
\begin{abstract}
The invariant theory of Killing tensors (ITKT) is extended by introducing the new concepts
of covariants and joint invariants of (product) vector spaces of Killing tensors defined
in pseudo-Riemannian spaces of constant curvature. The covariants are employed  to solve the problem of
 classification of the orthogonal
coordinate webs generated by non-trivial Killing tensors of valence two 
 defined in the Euclidean and Minkowski planes. Illustrative examples are provided. 
\end{abstract}

\bigskip 

\noindent Indexing Codes: 02.30Ik, 02.40.Ky, 02.20.Hj

\newpage

\section{Introduction}

The second half of the 19th century saw the development of the post-``Theorema
Egregium of Gauss'' differential geometry  going in two major directions.
Thus, Bernhard Riemann \cite{Ri} generalized Gauss' geometry of surfaces in the Euclidean
space by introducing the concept of a differentiable manifold of arbitrary
dimension and defining the inner product in terms of the metric tensor on 
the  spaces of tangent vectors. This remarkable work has evolved in time
into what is known today as (Riemannian) differential geometry. 
The other direction originated in  the celebrated   ``Erlangen Program'' of Felix Klein \cite{FK72, FK93}. 
According to his manifesto any branch of geometry can be interpreted as an invariant
theory with respect to a specific transformation group. Moreover, the main goal of any geometry is
the determination of those properties of geometrical figures that remain unchanged under the action
of  a transformation group. One of the main contributions of \'{E}lie Cartan to differential geometry,
in particular with his moving frames method \cite{Car}, is the blending of these two directions into
a single theory. An excellent exposition of this fact can be found in Sharpe \cite{SHA} (see also, for example, 
Arvanitoyeorgos \cite{Arv}). The
following diagram presented in \cite{SHA} elucidates the relationship among the different approaches
to geometry described above.

\begin{equation}
\begin{array}{ccc}
\mbox{Euclidean Geometry} & \stackrel{\mbox{\small generalization}}{\longrightarrow} & \mbox{Klein Geometries} \\[0.5cm]
 \downarrow\mbox{\small generalization} & & \mbox{\small generalization}\downarrow \\[0.5cm]
 \mbox{Riemannian Geometry} & \stackrel{\mbox{\small generalization}}{\longrightarrow} & \mbox{Cartan Geometries}
 \end{array}
 \label{Dia}
\end{equation}

Being a result of
the natural fusion of classical invariant theory (CIT) and the (geometric) study of Killing tensors
defined in pseudo-Riemannian manifolds of constant curvature, the  invariant theory  of
 Killing tensors (ITKT) formed recently a new direction of  research 
 \cite{HMS, MMS, JY, MST2, DHMS, MST3, MST4, MST5, MST6, MST1}, which, in view of the above, can be 
 rightfully placed into the theory initiated by Cartan. This is especially evident in the study of vector spaces of Killing tensors of valence two.
Indeed, by now a number of vector spaces of Killing tensors have been investigated
from this viewpoint by means of determining the corresponding sets of
fundamental {\em invariants} and, much like in CIT, using
them to solve the problem of equivalence  in each case. These results have been
employed  in applications arising in the  {\em theory of orthogonal
coordinate webs}  \cite{Bo, Dar, Ei1, Ei2, Ole,  Be, Kal, Ka,  Mi,  MST1, MST3, HMS}, where Killing tensors of valence two play a pivotal
role (see the review \cite{Be} for a complete list of references). Admittedly, an orthogonal coordinate web is an integral part of the geometry of the underlying
pseudo-Riemmanian manifold. Therefore the problem of group invariant classification of orthogonal coordinate webs in a specific
pseudo-Riemannian space of constant curvature is a problem of Felix Klein's approach to geometry, as well as that of Riemann, both leading to
the theory due to Cartan (see the diagram (\ref{Dia})). 

The main goal of this paper is to further the development of the invariant 
theory of Killing tensors by introducing the concepts of a {\em covariant} and 
a {\em joint invariant}. 
In  this setting they can be introduced  by establishing  a natural extension of the main ideas of 
 CIT to the geometric study of Killing tensors in pseudo-Riemannian geometry.
Furthermore, we  employ the latest generalization of  Cartan's method
of moving frames  due to Fels and Olver \cite{FO1,FO2} 
(see also \cite{Car,Gu,Gr,Gre, Kam, Ko} for more details and references) to determine complete systems of fundamental covariants
for the vector spaces of Killing tensors of valence two defined in the Euclidean and
Minkowski planes. The covariants are employed to classify in both cases
orthogonal coordinate webs generated by  Killing tensors. 
We also compare the results with the classifications of the orthogonal webs defined in the Minkowski plane obtained in McLenaghan {\em et al} \cite{MST3, MST6}
 by means of invariants only.

\section{Invariant theory of Killing tensors (ITKT) }

In this section we establish the requisite language and recall the basic notions of the invariant
theory of  Killing tensors (ITKT) defined in pseudo-Riemannian spaces of constant curvature. More specifically, 
we review what is known about ismetry group invariants  and extend the theory by introducing the concepts of
{\em covariants} and {\em joint invariants} of product vector spaces of Killing tensors in ITKT. 
Let $(M, {\bf g})$ be a pseudo-Riemannian manifold, $\dim\, M = n$. 
\begin{defi}
\label{DKT}
A {\em Killing tensor ${\bf K}$ of valence $p$ defined in $(M,{\bf g})$}

is a symmetric $(p,0)$ tensor satisfying the Killing tensor equation
\begin{equation}
\label{KE}
[{\bf K},{\bf g}]=0,
\end{equation}
where $[$ , $]$ denotes the Schouten bracket \cite{Sch}. When $p=1$, ${\bf K}$ is said to be a {\em Killing
vector (infinitesimal isometry)} and the equation (\ref{KE}) reads
$$
{\mathcal L}_{{\bf K}}{\bf g} = 0,
$$
where $\mathcal L$ denotes the  Lie derivative operator.
\end{defi}
\begin{rem}  {\rm 
Throughout this paper, unless otherwise specified, 
 $[$ , $]$ denotes the Schouten bracket, which is a generalization of  the usual Lie
 bracket of vector fields. 
}  \end{rem}  

Killing tensors appear naturally in many problems of classical mechanics,
general relativity, field theory and other areas. To demonstrate
this fact, let us consider the following example.

\begin{exa}  {\rm  
Let $({\bf X}_H, {\bf P}_0, H)$ be a Hamiltonian system defined  on
 $(M, {\bf g})$ by a natural Hamiltonian $H$ of the form  
\begin{equation}  
H ({\bf q}, {\bf p}) = \frac{1}{2}g^{ij}p_ip_j + V({\bf q}), 
\quad i,j  = 1,\ldots, n, \label{H}  
\end{equation}  
where $g^{ij}$ are the contravariant components of the corresponding  
metric tensor $\bf g$,  $({\bf q},{\bf p}) \in T^*M$ are the canonical 
 position-momenta coordinates and the Hamiltonian vector field  
${\bf X}_H$ is given by  
\begin{equation}  
{\bf X}_H = [{\bf P}_0, H] \label{X}  
\end{equation} 
 with respect to the canonical Poisson bi-vector   
${\bf P}_0 = \sum_{i=1}^n\partial/\partial q^i\wedge \partial/\partial p_i.$  
Assume also that the Hamiltonian system defined by (\ref{H})  admits a first integral
 of motion $F$ which is a polynomial function  of degree $m$ in the momenta:  
\begin{equation} 
\label{FI} 
F({\bf q}, {\bf p})=
 K^{i_1i_2\ldots i_m}({\bf q})p_{i_1}p_{i_2}\ldots p_{i_m}  + U({\bf q}),  
\end{equation}  
where $ 1\le i_1, \ldots, i_m \le n.$  Since the functions $H$ and $F$ are in involution, 
 the vanishing of the  Poisson bracket defined by 
${\bf P}_0$: 
\begin{equation}
\label{PB}
\{H, F\}_0 = {\bf P}_0\d H\d F = [[{\bf P}_0, H],F]= 0
\end{equation}
  yields 
\begin{equation} 
[{\bf K}, {\bf g}] =0,  
\quad \mbox{(Killing tensor equation)} \label{KT1}  
\end{equation}  and  
\begin{equation}  
K^{i_1 i_2\ldots i_m}\frac{\partial V}{\partial q^{i_1}}p_{i_2}\ldots p_{i_m}  
= g^{ij}\frac{\partial U}{\partial q^i}p_j,  
\quad \mbox{(compatibility condition)}, \label{CC}  
\end{equation} 
 where the symmetric $(n,0)$-tensor $\bf K$ has the components  
$K^{i_1 i_2\ldots i_m}$ and  $1 \le $  $i,$ $j,$ $i_1,$ $\ldots,$ $ i_m$  
$ \le n$. Clearly, in view of Definition \ref{DKT} the equation  
(\ref{KT1}) confirms that $\bf K$ is a Killing tensor.  Furthermore,
 in the case $m=2$ (see Benenti \cite{Be})  the compatibility condition (\ref{CC})
 reduces to  ${\bf K}\d V = {\bf g} \d U$ or $\d (\hat{\bf K}\d V) = 0$,  
where the $(1,1)$-tensor $\hat{\bf K}$ is  given by  
$\hat{\bf K} = {\bf K}{\bf g}^{-1}$. We also note that the vanishing of the Poisson bracket 
(\ref{PB}) and the assumed form of the first integral $F$ (\ref{FI}) imply the following additional conditions:
$$\partial_iU = 0, \quad K^{i_1i_2\ldots i_m}\partial_{i_1}V = 0.$$
Indeed, the RHS of (\ref{FI}) does not have the terms which are polynomials of $\bf p$ of degrees less than $m$.
} \label{E1}  \end{exa}

In view of linear properties of the Schouten bracket the sets of Killing tensors of the same
valence form vector spaces in $(M, {\bf g})$. Let ${\cal K}^p(M)$ denote the vector space of Killing tensors of
valence $p\ge 1$ defined in $(M, {\bf g})$.  Assume also $\dim\, M = n$. Then if $(M, {\bf g})$ is 
a pseudo-Riemannian space of constant curvature, the dimension $d$ of the corresponding vector space ${\cal K}^p(M)$
for a given $p \ge 1$ is determined  by the {\em Delong-Takeuchi-Thompson (DTT) formula} \cite{De, Tak, Th} 
\begin{equation}
d = \dim\,{\mathcal K}^p(M) = \frac{1}{n}{n+p \choose p+1}{n+p-1 \choose p}, \quad p \ge 1.
\label{DTT}
\end{equation} 
That being the case, a Killing tensor of valence  $p\ge 1$  defined in a pseudo-Riemannian space 
$(M, {\bf g})$ of constant curvature can be viewed as an algebraic object,  or, an element 
of ${\cal K}^p(M)$. Note the vector space ${\cal K}^p(M)$ for a fixed $p \ge 1$ is  determined by  $d$ arbitrary parameters $(\alpha_1, \ldots, \alpha_d)$, where 
$d = \dim\,{\mathcal K}^p(M)$ is given by (\ref{DTT}).   This approach to the study of Killing tensors  introduced in \cite{MST6} differs 
significantly from the more conventional approach based on the property that Killing tensors defined in pseudo-Riemannian
spaces of constant curvature are sums of symmetrized tensor products of Killing vectors (see, for example, \cite{Th}). Moreover, 
the idea leads to a natural link between the study of vector spaces of Killing tensors and the classical theory of
invariants of vector spaces of  homogeneous polynomials, which 
has become in the last decade a growth industry once again (see Olver \cite{Olv} and
the references therein). Thus, it has been shown in a series of recent papers \cite{DHMS,MST1,MST2,MST3,MST4,MST5,MST6} that one can utilize the basic ideas
of classical invariant theory in the study of Killing tensors defined in pseudo-Riemannian spaces of constant curvature. The concept of
an {\em invariant} of ${\cal K}^p(M)$ was introduced in \cite{MST1} in the study of non-trivial Killing tensors of the vector space ${\cal K}^2(\mathbb{R}^2)$
generating orthogonal coordinate webs in the Euclidean plane. 

\subsection{Invariants}

It has been shown that  one can  determine the action of the isometry group $I(M)$ in the $d$-dimensional  space $\Sigma \simeq \mathbb{R}^d$ defined by 
 the parameters $\alpha_1,\ldots, \alpha_d$. In this view, the action is induced by the corresponding action of $I(M)$ in ${\cal K}^p(M)$, which, 
 in turn, is induced by the action of $I(M)$ in $M$.  More specifically, it induces the corresponding transformation laws for the parameters 
 $(\alpha_1, \ldots, \alpha_d)$ given by
\begin{equation}
\begin{array}{c}
\tilde{\alpha}_1 = \tilde{\alpha}_1(\alpha_1,\ldots, \alpha_d, g_1, \ldots, g_r), \\
\tilde{\alpha}_2 = \tilde{\alpha}_2(\alpha_1,\ldots, \alpha_d, g_1, \ldots, g_r), \\
\vdots \\
\tilde{\alpha}_d = \tilde{\alpha}_d(\alpha_1,\ldots, \alpha_d, g_1, \ldots, g_r), 
\end{array}
\label{TL}
\end{equation}
where $g_1, \ldots, g_r$ are local coordinates on $I(M)$ 
that parametrize the group and $r =  \dim\, I(M) = \frac{1}{2}n(n+1)$. The formulas (\ref{TL}) can be obtained in each case by making
use of the standard transfomation rules for tensor components. We note that the action of $I(M)$ can be considered in 
the spaces $M$ and $\Sigma$ concurrently, provided there is an isomorphism between the corresponding group actions (see below). 

\begin{defi} Let $(M,{\bf g})$  be a pseudo-Riemannian manifold of constant curvature.  For a fixed $p \ge 1$ consider the corresponding 
space ${\cal K}^p(M)$ of Killing tensors of valence $p$ defined in $(M,{\bf g})$. A smooth function 
${\cal I}:\, \Sigma \rightarrow \mathbb{R}$ defined in the space of functions on the parameter space $\Sigma$ is said
to be an $I(M)$-{\em invariant of the vector space ${\cal K}^p(M)$} iff it satisfies the condition
\begin{equation}
{\cal I} = F(\alpha_1, \ldots, \alpha_d) = F(\tilde{\alpha}_1, \ldots, \tilde{\alpha}_d)
\label{Inv}
\end{equation} 
under the transformation laws (\ref{TL}) induced by the isometry group $I(M)$. 
\end{defi} 

The main problem of  invariant theory is to describe the whole space of invariants (covariants, joint invariants) for a given vector space under the action of
a group. To solve this problem one has to 
find a set of {\em fundamental
invariants} ({\em covariants, joint invariants}) with the property that any other invariant (covariant, joint invariant) is a (analytic) function of 
 the fundamental invariants (covariants, joint invariants). The Fundamental Theorem
on Invariants of  a regular Lie group action \cite{Olv} determines the number of
fundamental invariants required to define the whole of the space of
$I(M)$-invariants: 
\begin{theo} Let $G$ be a Lie group acting regularly on an $m$-dimensional manifold $X$ with $s$-dimensional orbits. 
Then, in a neighbourhood $N$ of each point $x_0 \in X$, there exist $m-s$ functionally independent
$G$-invariants $\Delta_1, \ldots,$ $ \Delta_{m-s}$. Any other $G$-invariant $\cal I$ defined near $x_0$ can be locally
uniquely expressed as an analytic function of the fundamental invariants 
through ${\cal I} = F(\Delta_1,$ $ \ldots,$ $ \Delta_{m-s})$. 
\label{FT}
\end{theo}
Hence, if we assume that the group $I(M)$, $\dim\, I(M) = r= \frac{1}{2}n(n+1)$ acts in a subspace $\Sigma_r$ of the 
parameter space  $\Sigma$
defined by the corresponding ${\cal K}^p(M)$, $p \ge 1$ regularly with $r$-dimensional orbits, then,  according to Theorem \ref{FT}, the number of fundamental
invariants required to describe the whole space of $I(M)$-invariants of ${\cal K}^p(M)$ is $d-r$, where 
$d$ is given by (\ref{DTT}) (note $d \ge r$). This has been shown to be the case for the vector spaces 
${\cal K}^2(\mathbb{R}^2)$ \cite{MST1}, ${\cal K}^2(\mathbb{R}^2_1)$ \cite{MST3}, ${\cal K}^3(\mathbb{R}^2)$ \cite{MST2}  and ${\cal K}^2(\mathbb{R}^3)$ \cite{HMS}, where 
$\mathbb{R}^2$, $\mathbb{R}^2_1$ and $\mathbb{R}^3$ denote the Euclidean, Minkowski planes and the Euclidean space
respectively. The dimension of the orbits of the isometry group $I(M)$ acting in $\Sigma$ is not always the same as
 the dimension of the group. For example, this is the case  for the vector space ${\cal K}^1(\mathbb{R}^3)$ \cite{DHMS}. 
 To determine 
the dimension of the orbits one can use the infinitesimal generators of the group $I(M)$ in $\Sigma$. 

In what follows we use the approach introduced in \cite{MST6}. Let ${\bf X}_1, \ldots, {\bf X}_r \in {\cal X}(M)$ be 
the infinitesimal generators (Killing vector fields) of the Lie group $I(M)$ 
acting on $M$. Note $\mbox{Span}\, \{ {\bf X}_1, \ldots, {\bf X}_r\} = {\cal K}^1(M) = i(M)$, where $i(M)$ is the Lie algebra
of the Lie group $I(M)$. For a fixed $p\ge 1$, consider the corresponding vector space ${\cal K}^p(M)$. To determine 
the action of $I(M)$ in the space $\Sigma$, we find first the infinitesimal generators of $I(M)$ in $\Sigma$. Consider
$\Diff\, \Sigma$, it defines the corresponding space $\Diff\, {\cal K}^p(M)$, whose elements are determined by the elements
of $\Diff\, \Sigma$ in an obvious way. Let ${\bf K}^0 \in \Diff\, {\cal K}^p(M)$. Note ${\bf K}^0$ is determined by $d$ parameters
$\alpha_i^0(\alpha_1,\ldots, \alpha_d),$ $i=1,\ldots, d$, which are functions of  $\alpha_1, \ldots, \alpha_d$ - the 
parameters of $\Sigma$. Define now a map $\pi:\, \Diff \, {\cal K}^p(M) \rightarrow
{\cal X}(\Sigma),$  given by 
\begin{equation}
{\bf K}^0 \rightarrow \sum_{i=1}^d\alpha_i^0(\alpha_1,\ldots, \alpha_d)\frac{\partial}{\partial \alpha_i}.
\label{pi}
\end{equation}
To specify the action of $I(M)$ in $\Sigma$, we have to find the counterparts of the generators ${\bf X}_1, \ldots, {\bf X}_r$ in ${\cal X}(\Sigma)$.
Consider the composition $\pi\circ {\cal L}$, where $\pi$ is defined by (\ref{pi}) and ${\cal L}$ is the Lie derivative operator. Let 
${\bf K}$ be the general Killing tensor of ${\cal K}^p(M)$, in other words $\bf K$ is the general solution to the Killing tensor equation
(\ref{KE}). Note, for $p=2$ we have ${\bf K} = \mbox{Span}\, \{{\bf g}, {\bf K}_1, \ldots, {\bf K}_{d-1}\}$, where $ \{{\bf g}, {\bf K}_1, \ldots, {\bf K}_{d-1}\}$
is a basis of the vector space ${\cal K}^2(M)$ and $\bf g$ is the metric of $(M,{\bf g})$. Next, define
\begin{equation}
\label{comv1}
{\bf V}_i = \pi{\cal L}_{{\bf X}_i}\,{\bf K}, \quad i = 1, \ldots r, 
\end{equation}
The composition map $\pi \circ {\cal L}:\, i(M) \rightarrow {\cal X}(\Sigma)$ maps the generators ${\bf X}_1, \ldots, {\bf X}_r$ to
${\cal X}(\Sigma)$. 

\begin{con}\cite{MST2}
\label{Conj1}
Suppose the generators ${\bf X}_1, \ldots, {\bf X}_r$ of $i(M)$ satisfy the following
commutator relations:
\begin{equation}
\label{comr}
[{\bf X}_i, {\bf X}_j] = c^k_{ij}{\bf X}_k, \quad i,j,k = 1, \ldots, r,
\end{equation}
where $c^k_{ij}$, $i,j,k = 1, \ldots, r$ are the structural constants. Then the corresponding vector fields
${\bf V}_i \in {\cal X}(\Sigma),$  defined by (\ref{comv1}) satisfy the same commutator relations:
\begin{equation}
[{\bf V}_i, {\bf V}_j] = c^k_{ij}{\bf V}_k, \quad i,j,k = 1, \ldots, r.
\label{commutator}
\end{equation}
Therefore the map $F_* := \pi \circ {\cal L}: \, i(M) \rightarrow i_{\Sigma}(M)$  is a Lie algebra isomorphism, where
$i_{\Sigma}(M)$ is the Lie algebra generated by ${\bf V}_1, \ldots, {\bf V}_r$.
\end{con}

We emphasize that the technique of the 
Lie derivative  deformations used here is a very powerful tool. 
It was used before, for example,  in \cite{Sm} 
to generate compatible Poisson bi-vectors in  the theory of 
bi-Hamiltonian systems. The idea introduced in \cite{Sm}
 was  utilized in  \cite{Ser} and applied to a different class of 
integrable systems.  
The validity of the formula (\ref{commutator}) can  be confirmed directly on a case by case basis, provided that the
general form of a Killing tensor ${\bf K}^p \in {\cal K}^p(M)$ is available. 
The proof of the general  statement of Conjecture \ref{Conj1} will be 
published elsewhere \cite{MMS}.  
\begin{rem} {\rm Alternatively, the generators (\ref{comv1}) can be obtained from the formulas for the action of
the group (\ref{TL}) in the usual way taking into account that a Lie algebra is the tangent space at the unity of
the corresponding Lie group. We note, however, that in this way  the formulas (\ref{TL}) are not easy to derive in general. 
}

\end{rem}

In view of the isomorphism exhibited in the conjecture and the fact that  invariance of a function under an entire Lie group is
equivalent to the infinitesimal invariance under the infinitesimal generators of the corresponding Lie algebra one can  
determine  a set of fundamental invariants by solving the system of PDEs
\begin{equation}
\label{system}
{\bf V}_i(F) = 0, \quad i = 1, \ldots, r
\end{equation}
for an analytic function $F: \, \Sigma \rightarrow \mathbb{R}$, where the vector fields ${\bf V}_i,$ $i = 1, \ldots, r$ are the
generators defined by (\ref{comv1}). As is specified by Theorem \ref{FT}, the general solution to the system (\ref{system}) is an analytic function
$F$ of the fundamental invariants. The number of fundamental invariants is $d-s$, where $d$ is specified by the DTT-formula (\ref{DTT}) 
and $s$ is the dimension of the orbits of $I(M)$ acting regularly in the parameter space $\Sigma$. To determine $s$ and the subspaces of $\Sigma$
where the isometry group acts with orbits of the same dimension, one employs the result of the following proposition \cite{Olv}. 
\begin{prop}
Let a Lie group $G$ act on $X$, {\gothic g} is the corresponding Lie algebra and let $x\in X$. The vector space ${ S}|_x = \mbox{\em Span}\{{\bf V}_i(x)|\, {\bf V_i} \in \mbox{\gothic g}\}$ 
spanned by all vector fields determined by the infinitesimal generators at $x$ coincides with the tangent space  to the orbit 
${\cal O}_x$ of $G$ that passes through $x$, so $S|_x = T{\cal  O}_x|_x$. In particular, the dimension of ${\cal O}_x$ equals the 
dimension of $S|_x$. Moreover, the isotropy subgroup $G_x \subset G$ has dimension $\dim G - \dim {\cal O}_x = r-s.$
\label{Prop1}
\end{prop}

\begin{exa} {\rm Consider the action of the isometry group $I(\mathbb{R}_1^2)$ on the vector space ${\cal K}^2(\mathbb{R}_1^2)$.  More information
about the geometry of Minkowski plane $\mathbb{R}_1^2$ can be found in the monograph Thompson \cite{AT}.  The  general
form of the elements of ${\cal K}^2(\mathbb{R}_1^2)$ in terms of the standard pseudo-Cartesian coordinates $(t,x)$ is given by
\begin{equation}
\begin{array}{rcl}
{\bf K} & = & \displaystyle (\alpha_1 + 2\alpha_4x + \alpha_6x^2)\frac{\partial}{\partial t}
\odot  \frac{\partial}{\partial t} \\ [0.3cm]
& & + \displaystyle (\alpha_3 + \alpha_4t +\alpha_5x  +  \alpha_6tx)\frac{\partial}{\partial t}\odot
\frac{\partial}{\partial x} \\ [0.3cm]
& & + \displaystyle (\alpha_2 + 2\alpha_5t+\alpha_6t^2) \frac{\partial}{\partial x}\odot
\frac{\partial}{\partial x},
\end{array}
\label{gKT}
\end{equation}
The isometry group $I(\mathbb{R}_1^2)$ acts in the Minkowski
plane $\mathbb{R}_1^2$ parametrized by  $(t,x)$ as follows. 
\begin{equation}
     \left( \begin{array}{c} \tilde{t}\\ \tilde{x} \end{array} \right) 
     = \left( \begin{array}{cc}
        \cosh\phi & \sinh\phi\\
        \sinh\phi & \cosh\phi
        \end{array} \right) 
        \left( \begin{array}{c} t\\ x\end{array} \right)
        +  \left( \begin{array}{c} a\\b \end{array}\right),
 \label{RM}
 \end{equation}
where $\phi, a, b \in \mathbb{R}$ are local coordinates that parametrize the group $I(\mathbb{R}_1^2)$.
The generators of the Lie algebra $i(\mathbb{R}_1^2)$ of the isometry group with respect to the coordinates $(t,x)$ take the
following form:
\begin{equation} {\bf T} = \frac{\partial}{\partial t}, 
\quad {\bf X} = \frac{\partial}{\partial x}, 
\quad {\bf H} = x\frac{\partial}{\partial t} + t\frac{\partial}{\partial x} 
\label{TXH} 
\end{equation} 
corresponding to $t$- and $x$-translations and (hyperbolic) rotation, 
given with respect to the standard pseudo-Cartesian coordinates $(t,x)$. 
Note the generators (\ref{TXH}) of the Lie algebra $i(\mathbb{R}_1^2)$ 
enjoy the following commutator relations:  
\begin{equation}  \label{COMM}  
[{\bf T},{\bf X}] = 0, \quad 
[{\bf T}, {\bf H}] = {\bf X}, \quad 
[{\bf X}, {\bf H}] = {\bf T}.  
\end{equation} 
We use the formula (\ref{RM}) and  the transformation laws for the components of 
$(2,0)$ tensors  
\begin{equation} 
\tilde{K}^{ij}(\tilde{y^1},\tilde{y^2},\tilde{\alpha}_1,\ldots, \tilde{\alpha}_6)  
= K^{k\ell}(y^1,y^2,\alpha_1,\ldots, \alpha_6)\frac{\partial  
\tilde{y}^i}{\partial y^k}\frac{\partial  \tilde{y}^j}{\partial y^{\ell}}, 
\quad i,j, k,\ell = 1, 2, \label{KTR} 
\end{equation}  
where the tensor components $K^{ij}$ are given by (\ref{gKT}), 
$y^1 = t, y^2 = x.$
In view of (\ref{gKT}),
(\ref{RM}) and (\ref{KTR}) the transformation laws  (\ref{TL}) for the parameters $\alpha_i,$ $i=1,\ldots, 6$ 
take in this case the following form (see also \cite{Kal, MST3}).
\begin{equation}
 \label{transf}
 \begin{array}{rcl}
 \tilde{\alpha}_1 & = & \alpha_1\cosh^2\phi + 2\alpha_3\cosh\phi\sinh\phi + \alpha_2\sinh^2\phi + \alpha_6b^2\\[0.3cm]
  & & -2 (\alpha_4\cosh \phi  + \alpha_5\sinh \phi)b,\\[0.3cm]
  \tilde{\alpha}_2 & = & \alpha_1\sinh^2\phi + 2\alpha_3\cosh\phi\sinh\phi + \alpha_2\cosh^2\phi +
\alpha_6 a^2\\[0.3cm]
  & & -2 (\alpha_5\cosh \phi  + \alpha_4\sinh \phi)a,\\[0.3cm]
  \tilde{\alpha}_3&=& \alpha_3(\cosh^2\phi + \sinh^2\phi)
  +(\alpha_1 + \alpha_2)\cosh\phi\sinh\phi  \\[0.3cm]
 & &  - (a \alpha_4 + b \alpha_5)\cosh\phi - (a \alpha_5 + b \alpha_4)\sinh\phi + \alpha_6ab,\\[0.3cm]
 \tilde{\alpha}_4 & = & \alpha_4\cosh\phi + \alpha_5\sinh\phi - \alpha_6 b, \\[0.3cm]
 \tilde{\alpha}_5& = & \alpha_4\sinh \phi + \alpha_5 \cosh\phi -\alpha_6 a, \\[0.3cm]
 \tilde{\alpha}_6 &=& \alpha_6.
 \end{array}
 \end{equation}
}
\end{exa}

We note that the corresponding transformation formulas for the parameters obtained in \cite{MST3} 
were derived for  {\em covariant} Killing tensors. Accordingly, they differ somewhat from  (\ref{transf}) presented above
(compare with (7.6) in \cite{MST3}).
According to Proposition \ref{Prop1}, in order to determine the subspaces of $\Sigma$ where the orbits have the same dimensions, one has to check
the subspaces of $\Sigma$ where the system (\ref{system}) retains its rank. In many cases the system of PDEs (\ref{system}) can be solved 
by the method of characteristics. The determination 
of fundamental invariants by solving (\ref{system}) is the key idea  used in \cite{MST6} to adapt
 the {\em method of infinitesimal generators}  to
the problem of finding fundamental invariants of Killing tensors under the action of the isometry group. When the method of
characteristic fails, one can employ the {\em method of undetermined coefficients}  to find a set of fundamental invariants \cite{DHMS,HMS}. Alternatively, 
a set of fundamental invariants can be determined by using  the {\em  method of
moving frames} (see Section 3 for more details). To determine
the space of $I(\mathbb{R}_1^2)$-invariants, we employ the
procedure described above and derive the corresponding infinitesimal generators ${\bf V}_i$, $i=1,2,3$ by the formula (\ref{comv1}):
\begin{equation} \label{vecM}  
\begin{array}{rcl} 
{\bf V}_1 &=& \displaystyle \alpha_4\frac{\partial}{\partial {\alpha_3}} +2\alpha_5\frac{\partial}{\partial {\alpha_2}}+
\alpha_6 \frac{\partial}{\partial {\alpha_5}}, \\[0.3cm] 
 {\bf V}_2 &=& \displaystyle \alpha_5\frac{\partial}{\partial  {\alpha_3}}+2\alpha_4\frac{\partial}{\partial {\alpha_1}}+
 \alpha_6\frac{\partial}{ \partial {\alpha_4}},\\[0.3cm] 
 {\bf V}_3 &=& \displaystyle -2\alpha_3\frac{\partial}{\partial  {\alpha_1}}-\alpha_5\frac{\partial}{\partial  {\alpha_4}} 
 -(\alpha_1+\alpha_2)\frac{\partial}{\partial  {\alpha_3}}-2\alpha_3 \frac{\partial}{\partial {\alpha_2}}- \alpha_4\frac{\partial}{\partial {\alpha_5}}. 
\end{array}  
\end{equation} 
and then solve by the method of characteristic the corresponding system of PDEs (\ref{system}) with respect to (\ref{vecM}). Note the vector fields $-{\bf V}_i$, $i=1,2,3$
satysfy the same communator relations as (\ref{TXH}) (see (\ref{COMM})), which confirms Conjecture \ref{Conj1}. Ultimately, this leads to the following theorem. 
\begin{theo}
\label{TFM}
Any algebraic $I(\mathbb{R}^2_1)$-invariant ${ I}$ of the subspace of the parameter space $\Sigma$ of ${\cal K}^2(\mathbb{R}_1^2)$ defined 
by the condition that the vector fields (\ref{vecM}) are linearly independent 
can be (locally) uniquely expressed as an analytic function 
$${I}= F({\cal I}_1, {\cal I}_2, {\cal I}_3)$$
where the funcamental invariants ${\cal I}_i,$ $i=1,2,3$ are given by
\begin{equation}
\begin{array}{rcl}

{\cal I}_1 & = & (\alpha_4^2+\alpha_5^2-\alpha_6(\alpha_1+\alpha_2))^2 -4(\alpha_3\alpha_6-\alpha_4\alpha_5)^2, \\ [0.3cm]

{\cal I}_2 & = & \alpha_6(\alpha_1-\alpha_2) -  \alpha_4^2+\alpha_5^2,  \\ [0.3cm]

{\cal I}_3 & = & \alpha_6.

\end{array}
\end{equation}

\end{theo} 
The fact that ${\cal I}_3 = \alpha_6$ is a fundamental $I(\mathbb{R}_1^2)$-invariant of the vector space ${\cal K}^2(\mathbb{R}_1^2)$
trivially follows from the transformation formulas (\ref{transf}). The
fundamental $I(\mathbb{R}_1^2)$-invariant ${\cal I}_1$
was derived in \cite{MST3, MST6} in the study of the five-dimensional subspace of non-trivial Killing tensors of ${\cal K}^2(\mathbb{R}_1^2)$. 
As expected, in this case by Theorem \ref{FT}, we have  obtained 6 (dimension of the space) - 3 (dimension of  the orbits) = 3 
fundamental $I(\mathbb{R}_1^2)$-invariants of the vector space ${\cal K}^2(\mathbb{R}_1^2)$.

\subsection{Covariants}

Consider now the action of the isometry group $I(M)$ on the product space ${\cal K}^p(M)\times M,$ $ p\ge 1.$ As above
it induces the transformation laws on the {\em extended parameter space} $\Sigma \times M$, where $\Sigma$ is the parameter space of 
the vector space ${\cal K}^p(M)$: 
\begin{equation}
\begin{array}{c}
\tilde{\alpha}_1 = \tilde{\alpha}_1(\alpha_1,\ldots, \alpha_d, g_1, \ldots, g_r), \\
\tilde{\alpha}_2 = \tilde{\alpha}_2(\alpha_1,\ldots, \alpha_d, g_1, \ldots, g_r), \\
\vdots \\
\tilde{\alpha}_d = \tilde{\alpha}_d(\alpha_1,\ldots, \alpha_d, g_1, \ldots, g_r), \\
\tilde{x}_1 = \tilde{x}_1(x_1, \ldots, x_n, g_1, \ldots, g_r),\\
\tilde{x}_2 = \tilde{x}_2(x_1, \ldots, x_n, g_1, \ldots, g_r), \\ 
\vdots \\ 
\tilde{x}_n = \tilde{x}_n( x_1, \ldots, x_n, g_1, \ldots, g_r), 
\end{array}
\label{TLC}
\end{equation}
where as before $\alpha_1, \ldots, \alpha_d$ are the parameters of ${\cal K}^p(M)$ that define $\Sigma$, 
$g_1, \ldots, g_r$, $r = \frac{1}{2}n(n+1)$ are local parameters parametrizing the group $I(M)$ and $x_1, \ldots, x_n$ are local 
coordinates on the manifold $M$. 

\begin{defi}
An {\em  $I(M)$-covariant} of the vector space ${\cal K}^p(M)$ $p \ge 1$
is a  function $C: \, \Sigma \times M \rightarrow \mathbb{R}$ satisfying the condition
\begin{equation}
C = F(\alpha_1, \ldots, \alpha_d,x_1, \ldots, x_n) = F
(\tilde{\alpha}_1, \ldots, \tilde{\alpha}_d,\tilde{x}_1 \ldots, \tilde{x}_n)
\end{equation}
under the transformation laws (\ref{TLC}) induced by the isometry group $I(M)$,
where $\Sigma$ is the parameter space of ${\cal K}^p(M)$.
\end{defi}
Conjecture \ref{Conj1} entails the following corollary.
\begin{coro}
\label{CoroCov}
Consider the product vector space ${\cal K}^p(M)\times M,$ $ p\ge 1.$
 Define the vector fields
\begin{equation}
{\bf V}'_i := {\bf V}_i + {\bf X}_i \quad i = 1, \ldots, r,
\label{tVC}
\end{equation}
where ${\bf V}_i,$ $i = 1, \ldots,r$ are
the  infinitesimal generators of the Lie algebra $i(M)$ in the parameter space $\Sigma$
of the vector space ${\cal K}^p(M)$  obtained via (\ref{comv1}) and ${\bf X}_i,$ $i =1, \ldots, r$ are the generators of
$i(M)$. Then
the vector fields ${\bf V}'_1, \ldots, {\bf V}'_r$ enjoy the same commutator relations as the generators 
${\bf X}_1, \ldots, {\bf X}_r$ of $i(M)$ in ${\cal X}(M)$: 
\begin{equation}
[{\bf V}'_i, {\bf V}'_j] = c^k_{ij}{\bf V}'_k, \quad i,j,k = 1, \ldots, r,
\label{comVC}
\end{equation}
where the structural constants $c_{ij}^k$ are as in (\ref{comr}). 
\end{coro}
\noindent {\em Proof.} Straightforward. \hfill $\Box$

\smallskip

Therefore, in view of the above, $I(M)$-covariants of a vector space ${\cal K}^p(M)$ can be obtained by solving the corresponding
system of PDEs generated by the vector fields (\ref{tVC}):
\begin{equation}
{\bf V}'_i (F) = 0, \quad i = 1, \ldots, r.
\end{equation}
Alternatively, one can employ the method of moving frames. To demonstrate how it works
in the framework of ITKT  we shall employ the method
in Section 3 to compute the covariants of the vector spaces ${\cal K}^2(\mathbb{R}^2)$ and ${\cal K}^2(\mathbb{R}_1^2)$.

\subsection{Joint invariants}

Consider now the action of the isometry group $I(M)$ on the product space ${\cal K}^{\ell}(M) \times {\cal K}^m(M)\times \cdots \times {\cal K}^q(M),$
$\ell, m, \ldots, q \ge 1$. Let $\alpha_1, \ldots, \alpha_d$, $\beta_1, \ldots, \beta_e$, $\ldots$, $\gamma_1,\ldots, \gamma_f$ be the parameters of
the vector spaces ${\cal K}^{\ell}(M)$, ${\cal K}^m(M),$ $\ldots,$ ${\cal K}^q(M)$ respectively, where $d, e, \ldots, f$ are the corresponding
dimensions determined by (\ref{DTT}). Then the action of the isometry group $I(M)$ induces the corresponding transformation laws for
the parameters $\alpha_1, \ldots, \alpha_d$, $\beta_1, \ldots, \beta_e$, $\ldots$, $\gamma_1,\ldots, \gamma_f$: 
\begin{equation}
\begin{array}{c}
\tilde{\alpha}_1 = \tilde{\alpha}_1(\alpha_1,\ldots, \alpha_d, g_1, \ldots, g_r), \\
\tilde{\alpha}_2 = \tilde{\alpha}_2(\alpha_1,\ldots, \alpha_d, g_1, \ldots, g_r), \\
\vdots \\
\tilde{\alpha}_d = \tilde{\alpha}_d(\alpha_1,\ldots, \alpha_d, g_1, \ldots, g_r), \\
\tilde{\beta}_1 = \tilde{\beta}_1(\beta_1,\ldots, \beta_e, g_1, \ldots, g_r), \\
\tilde{\beta}_2 = \tilde{\beta}_2(\beta_1,\ldots, \beta_e, g_1, \ldots, g_r), \\
\vdots \\
\tilde{\beta}_e = \tilde{\beta}_e(\beta_1,\ldots, \beta_e, g_1, \ldots, g_r), \\
\vdots\\
\tilde{\gamma}_1 = \tilde{\gamma}_1(\gamma_1,\ldots, \gamma_f, g_1, \ldots, g_r), \\
\tilde{\gamma}_2 = \tilde{\gamma}_2(\gamma_1,\ldots, \gamma_f, g_1, \ldots, g_r), \\
\vdots \\
\tilde{\gamma}_f = \tilde{\gamma}_f(\gamma_1,\ldots, \gamma_f, g_1, \ldots, g_r), 
\end{array}
\label{TLP}
\end{equation}
where as before $g_1, \ldots, g_r$ are local coordinates on $I(M)$ 
that parametrize the group and $r =  \dim\, I(M) = \frac{1}{2}n(n+1)$.
This observation leads us to introduce the concept of a {\em joint $I(M)$-invariant}. 
\begin{defi}
A {\em joint $I(M)$-invariant} of the product space ${\cal K}^{\ell}(M) \times$ ${\cal K}^m(M)$ $\times \cdots$ $\times {\cal K}^q(M),$
is a  function $J: \, \Sigma^{\ell}\times \Sigma^m \times \cdots \times \Sigma^q \rightarrow \mathbb{R}$ satisfying the condition
\begin{equation}
\begin{array}{rcl}
J & = & F(\alpha_1, \ldots, \alpha_d,\beta_1 \ldots, \beta_e,\ldots, \gamma_1\ldots, \gamma_f) \\[0.3cm]
 & = & F
(\tilde{\alpha}_1, \ldots, \tilde{\alpha}_d,\tilde{\beta}_1 \ldots, \tilde{\beta}_e,\ldots, \tilde{\gamma}_1\ldots, \tilde{\gamma}_f) 
\end{array}
\end{equation}
under the transformation laws (\ref{TLP}) induced by the isometry group $I(M)$.
\end{defi}

In this case again Conjecture \ref{Conj1} entails the following corollary.
\begin{coro}
\label{Coro1}
Consider the product vector space 
\begin{equation}
\label{ps}
{\cal K} = {\cal K}^{\ell}(M) \times {\cal K}^m(M)\times \cdots \times {\cal K}^q(M),
\end{equation}
where $\ell, m, \ldots, q \ge 1$. Define the vector fields
\begin{equation}
\tilde{\bf V}_i := {\bf V}_i^{\ell} + {\bf V}_i^m + \cdots + {\bf V}_i^q, \quad i = 1, \ldots, r,
\label{tV}
\end{equation}
where $\{{\bf V}_i^{\ell}\}, \{{\bf V}_i^m\},\ldots,  \{{\bf V}_i^q\}$, $i = 1, \ldots, r$ are
the sets of infinitesimal generators of the Lie algebra $i(M)$ in the parameter spaces $\Sigma^{\ell}$, $\Sigma^m, \ldots, \Sigma^q$
of the vector spaces ${\cal K}^{\ell}(M), {\cal K}^q(M), \ldots,  {\cal K}^n(M)$ respectively obtained via (\ref{comv1}). Then
the vector fields $\tilde{\bf V}_1, \ldots, \tilde{\bf V}_r$ enjoy the same commutator relations as the generators 
${\bf X}_1, \ldots, {\bf X}_r$ of $i(M)$ in ${\cal X}(M)$: 
\begin{equation}
[\tilde{\bf V}_i, \tilde{\bf V}_j] = c^k_{ij}\tilde{\bf V}_k, \quad i,j,k = 1, \ldots, r,
\label{comVt}
\end{equation}
where the structural constants $c_{ij}^k$ are as in (\ref{comr}). 
\end{coro}
\noindent {\em Proof.} Straightforward. \hfill $\Box$

\smallskip

\begin{exa}{\rm
Consider the product vector space ${\cal K}^1(\mathbb{R}^2) \times {\cal K}^2(\mathbb{R}^2)$. The general form of the 
elements of ${\cal K}^1(\mathbb{R}^2)$ (Killing vectors) with respect to the Cartesian coordinates is given by
\begin{equation} 
{\bf K}^1 = (\alpha_1 + \alpha_3y)\frac{\partial}{\partial x} + (\alpha_2 - \alpha_3x)\frac{\partial}{\partial y},
\label{gKv}
\end{equation}
while the (contravariant) elements of ${\cal K}^2(\mathbb{R}^2)$ assume  the following general form with respect to the
same coordinate system:
\begin{equation}
\begin{array}{rcl}
{\bf K}^2 & = & \displaystyle (\beta_1 + 2\beta_4y + \beta_6y^2)\frac{\partial}{\partial x}
\odot  \frac{\partial}{\partial x} \\ [0.3cm]
& & + \displaystyle (\beta_3 - \beta_4x-\beta_5y - \beta_6xy)\frac{\partial}{\partial x}\odot
\frac{\partial}{\partial y} \\ [0.3cm]
& & + \displaystyle (\beta_2 + 2\beta_5x+\beta_6x^2) \frac{\partial}{\partial y}\odot
\frac{\partial}{\partial y},
\end{array}
\label{gKt}
\end{equation}
where $\odot$ denotes the symmetric tensor product. The formulas (\ref{gKv}) and (\ref{gKt}) put in evidence
that the corresponding parameter spaces $\Sigma^1$ and $\Sigma^2$ are determined by the three parameters 
$\alpha_i,$ $i= 1,\ldots, 3$ and the six parameters $\beta_i,$ $i = 1, \ldots, 6$ respectively. Let $I(\mathbb{R}^2)$ be the 
proper Euclidean group that consists of the orientation-preserving isometries of $\mathbb{R}^2$ (rigid motions). Its action
in $\mathbb{R}^2$ can be described as the semi-direct product of rotations and translations. In view of its standard parametrization,
we have the transformation of the Cartesian coordinates $ {\bf x} = (x,y)$ 
     \begin{equation}
     \label{tx}
     \tilde{\bf x} = R_{\theta}{\bf x} + {\bf a}, \quad  R_{\theta}
     = \begin{bmatrix}\cos\theta & -\sin\theta\\
\sin\theta & \;\;\,\cos\theta
 \end{bmatrix}
     \in SO(2), \quad  {\bf a} = (a,b) \in {\mathbb R}^2. 
     \end{equation}
Note, the generators of $i(\mathbb{R}^2) = {\cal K}^1(\mathbb{R}^2)$, which is the
Lie algebra of the Lie group $I(\mathbb{R}^2)$,  are given with 
respect to the Cartesian coordinates by
\begin{equation}
\label{gen}
{\bf X} = \frac{\partial}{\partial x}, \quad {\bf Y} = \frac{\partial}{\partial y}, \quad {\bf R} = x\frac{\partial}{\partial y} - y\frac{\partial}{\partial x},
\end{equation}
whose flows are translations and a rotation respectively. Employing the construction (\ref{comv1}), we derive two triples of the vector fields representing 
the generators (\ref{gen}) in ${\cal X}(\Sigma^1)$:
\begin{equation}
\label{gen1}
\begin{array}{l}
{\bf V}^1_1 = \displaystyle  -\alpha_3\frac{\partial}{\partial \alpha_2},\\ [0.3cm]
 {\bf V}^1_2 = \displaystyle  \alpha_3\frac{\partial}{\partial \alpha_1},\\ [0.3cm]
  {\bf V}^1_3 = \displaystyle \alpha_1\frac{\partial}{\partial \alpha_2} - \alpha_2\frac{\partial}{\partial \alpha_1}
\end{array} 
\end{equation} 
and ${\cal X}(\Sigma^2)$: 
\begin{equation}
\label{gen2}
\begin{array}{l}
{\bf V}^2_1 = \displaystyle -2\beta_5\frac{\partial}{\partial \beta_2} - \beta_4\frac{\partial}{\partial \beta_3} + \beta_6\frac{\partial}{\partial \beta_5},
\\ [0.3cm]
{\bf V}^2_2 = \displaystyle 2\beta_4\frac{\partial}{\partial \beta_1} - \beta_5\frac{\partial}{\partial \beta_3} + \beta_6\frac{\partial}{\partial \beta_6}, 
\\ [0.3cm] 
{\bf V}^2_3 = \displaystyle  -2\beta_3\Big(\frac{\partial}{\partial \beta_1} - \frac{\partial}{\partial \beta_2}\Big) + (\beta_1 - \beta_2)\frac{\partial}{\partial \beta_3}
+ \beta_5\frac{\partial}{\partial \beta_4} - \beta_4\frac{\partial}{\partial \beta_5}
\end{array}
\end{equation}
respectively. We note that in view of Conjecture \ref{Conj1} both the vector fields (\ref{gen1}) and the vector fields
(\ref{gen2}) satisfy the same  commutator relations as the generators of $i(\mathbb{R}^2)$ 
(\ref{gen}). By Corollary \ref{Coro1} this fact entails immediately
that the vector fields $\{\tilde{\bf V}_i\}$, $i=1,2,3$ defined by
\begin{equation}
\tilde{\bf V}_i := {\bf V}^1_i + {\bf V}_i^2, \quad i =1,2,3
\label{tVexa}
\end{equation} 
also enjoy the same commutator relations. This property can be also verified directly. 
Therefore we have determined the action of $I(\mathbb{R}^2)$ in the product space $\Sigma^1\times\Sigma^2$. To determine
the dimension of the orbits of the group we use the result of Propositon \ref{Prop1}. Thus, the orbits of the isometry
group $I(\mathbb{R}^2)$ acting in $\Sigma^1\times\Sigma^2$ are three-dimensional in the subspace  ${\cal S}_3 \subset \Sigma^1\times\Sigma^2$, 
where the generators (\ref{tVexa}) are linearly independent. According to Theorem \ref{FT},
The number of fundamental invariants in ${\cal S}_3$
is 9 (dimension of $\Sigma^1\times\Sigma^2$) - 3 (dimension of the orbits in ${\cal S}_3$) = 6.  Some of 
these fundamental invariants may be the 
fundamental invariants of the group action in the vector spaces ${\cal K}^1(\mathbb{R}^2)$ and ${\cal K}^2(\mathbb{R}^2)$. Indeed, it is 
instructive at this point to review the transformations imposed on the 9 parameters $(\alpha_1, \alpha_2, \alpha_3,$ $\beta_1,\beta_2,\beta_3,\beta_4,\beta_5, \beta_6)$
of the product space $\Sigma^1\times\Sigma^2$ by the group action:
\begin{equation}
\begin{array}{rcl}
\tilde{\alpha_1} & = & \alpha_1 \cos\theta  - \alpha_2\sin\theta - b\alpha_3, \\ [0.3cm]
\tilde{\alpha_2} & = & \alpha_1 \sin\theta + \alpha_2 \cos\theta +a\alpha_3, \\ [0.3cm]
\tilde{\alpha_3} & = & \alpha_3, \\[0.3cm] 
\tilde{\beta_1} &=& \beta_1\cos^2\theta - 2\beta_3\cos\theta\sin\theta + \beta_2\sin^2\theta - 2b\beta_4\cos\theta - 2b\beta_5\sin\theta
\\[0.3cm] 
                  & & + \beta_6b^2, \\ [0.3cm]
\tilde{\beta_2} & =& \beta_1\sin^2\theta - 2\beta_3\cos\theta\sin\theta +
\beta_2\cos^2\theta - 2a\beta_5\cos\theta + 2a\beta_4\sin\theta \\ [0.3cm]
 & & + \beta_6a^2,\\[0.3cm] 
\tilde{\beta_3} & = & (\beta_1-\beta_2)\sin\theta\cos\theta + \beta_3(\cos^2\theta - \sin^2\theta) 
+ (a\beta_4 + b\beta_5)\cos\theta \\ [0.3cm] 
 & &  + (a\beta_5 - b\beta_4)\sin\theta - \beta_6ab, \\[0.3cm]
 \tilde{\beta_4} & = & \beta_4\cos\theta + \beta_5\sin\theta - \beta_6b, \\[0.3cm] 
 \tilde{\beta_5} & = & \beta_5\cos\theta - \beta_4\sin\theta - \beta_6a,\\[0.3cm]
 \tilde{\beta_6} &=& \beta_6, 		  
\end{array}
\label{TL1}
\end{equation} 
where $(\theta, a, b)$ given by (\ref{tx}) parametrize the isometry group $I(\mathbb{R}^2)$. 
Hence, the dimension of the orbits in this subspace coincides with the dimension of the group.  We also observe that $\alpha_3$ and $\beta_6$ are
fundamental $I(\mathbb{R}^2)$-invariants of the group action in $\Sigma^1\times\Sigma^2$. }
\end{exa}

 To determine the remaining four fundamental invariants
we use the method of characteristics to  solve the system of linear PDEs
\begin{equation}
\label{tilV}
\tilde{\bf V}_i(F) = 0, \quad i=1,2,3
\end{equation}
where  $F:\, \Sigma^1\times\Sigma^2 \rightarrow \mathbb{R}$ and  the vector fields  $\tilde{\bf V}_i,\, i=1,2,3$ are given by
(\ref{tVexa}). 
Having solved  the system of PDEs (\ref{tilV}), we have therefore proven the following result. 
\begin{theo}
Any algebraic joint $I(\mathbb{R}^2)$-invariant $I$ defined over
 the subspace of $\Sigma^1\times\Sigma^2$
where the vector fields (\ref{tVexa}) are linearly independent can be locally uniquely expressed as an analytic function 
$$I= F({\cal I}_1, {\cal I}_2, {\cal I}_3, {\cal I}_4, {\cal J}_1, {\cal
J}_2),$$
where the fundamental joint $I(\mathbb{R}^2)$-invariants ${\cal I}_i, {\cal J}_j,\, i=1,\ldots, 4, j
=1,2$ are given by
\begin{equation}
\begin{array}{rcl}
{\cal I}_1 &=& [\beta_6(\beta_1-\beta_2) + \beta_5^2 - \beta_4^2]^2 +
4(\beta_3\beta_6 +\beta_4\beta_5)^2, \\[0.3cm]
{\cal I}_2 & = & \beta_6(\beta_1+\beta_2) - \beta_4^2 - \beta_5^2, \\[0.3cm]
{\cal I}_3 & = & \beta_6, \\[0.3cm]
{\cal I}_4 & = & \alpha_3, \\[0.3cm]
{\cal J}_1 & = &  (\beta_6\alpha_2+\beta_5\alpha_3)^2 + (\beta_6\alpha_1 - \beta_4\alpha_3)^2, \\[0.3cm]
{\cal J}_2 & = & (\beta_6\alpha_2 + \alpha_3\alpha_5)(\beta_6\beta_2 - \beta_5^2) + 
2(\beta_3\beta_6+\beta_4\beta_5)(\beta_6\alpha_1 - \beta_4\alpha_3).
\end{array} 
\end{equation}
\label{Th3}
\end{theo}
The fundamental joint $I(\mathbb{R}^2)$-invariants ${\cal I}_i, i=1,2,3$ are the fundamental $I(\mathbb{R}^2)$-invariants of
the vector space ${\cal K}^2(\mathbb{R}^2)$ (${\cal I}_1$ was derived in \cite{MST6}), while ${\cal I}_4$ is the fundamental  $I(\mathbb{R}^2)$-invariant
of the vector space ${\cal K}^1(\mathbb{R}^2)$. Note the fundamental $I(\mathbb{R}^2)$-invariants ${\cal J}_1$ and
${\cal J}_2$ are ``truly'' joint $I(\mathbb{R}^2)$-invairants of the vector spaces ${\cal K}^1(\mathbb{R}^2)$ and 
${\cal K}^2(\mathbb{R}^2)$. Therefore we have introduced an analogue of the concept of a joint invariant
in the classical invariant theory of homogeneous polynomials (refer to \cite{Hil} for more details). 
The problem of the determination of fundamental invariants, solved in this section for a particular (product) vector space of 
Killing tensors (Theorem \ref{Th3}) by the {\em method of infinitesimal generators},  can also be solved by the purely algebraic {\em method of moving frames}. 
This is the subject of the considerations that follow. 

\section{The method of moving frames}

The method of moving frames, indroduced originally by Cartan \cite{Car}, is a powerful technique that can be employed 
to solve a wide range of equivalence-type
problems. In its original interpretation it is based on an equivariant map from the space of submanifolds to a bundle of frames.
 The simplest example of a
moving frame
is the Frenet frame $\{{\bf t},{\bf n}\}$ of a regular curve $\gamma \in \mathbb{R}^2$ parametrized by its arc length. In this case the equivariant map
assigns to each point on the curve $\gamma (s)$ the corresponding frame $\{{\bf t}(s),{\bf n}(s)\}$. Clearly,  the moving frame along $\gamma$ can
be obtained from a fixed frame via a combination of rotations and/or translations. This puts in evidence that there is a natural isomorphism between
the moving frame and the orientation-preserving isometry group (Euclidean group) $I(\mathbb{R}^2)$. This is the essence of the later generalizations of the moving frame method \cite{Gu,
Gr, Gre}, where 
the moving frame was viewed as an equivariant map from the space of submanifolds to the group itself. In recent works by Fels and Olver \cite{FO1,FO2}
the classical moving frame method was further generalized to completely general transformation groups, including infinite-dimensional Lie pseudo-groups
(see also Kogan \cite{Ko}). Ultimately, the authors have succeeded in bringing the theory up to the level where the bundle of frames is no longer needed. 
We very briefly review the basic definitions and results  of the moving frames theory in its modern formulation (for a complete review, see \cite{Olv}). 
\begin{defi}
A {\em moving frame} is a smooth, $G$-equivariant map $\rho:\, M \rightarrow G$, 
where $G$ is a $r$-dimensional group acting smoothly on an $n$-dimensional  underlying manifold $M$.
\end{defi} 
\begin{theo} A moving frame exists in a neighborhood of a point ${\bf x} \in M$ iff $G$ acts freely and
regularly near $\bf x$. 
\end{theo}
To construct a moving frame, one employs  Cartan's {\em normalization method} \cite{Car}. 
\begin{theo}
Let $G$ act freely and regularly on $M$ and let $K \subset M$ be a (local) cross-section to the group orbits. Given
${\bf x} \in M$, let ${\bf g} = \rho ({\bf x})$ be the unique group element that maps $\bf x$ to the cross-section: 
${\bf g}\cdot {\bf x} = \rho({\bf x})\cdot {\bf x} \in K.$ Then $\rho: \, M \rightarrow G$ is a right moving frame. 
\end{theo}
More specifically, let ${\bf x} = (x_1,\ldots, x_n) \in M$ be local coordinates. Consider the explicit 
formulas for the coordinate transformations induced by the action of $G$: $\omega({\bf g},{\bf x}) = {\bf g}\cdot {\bf x}$. The 
right moving frame ${\bf g} = \rho({\bf x})$ can be constructed by making use of a {\em coordinate cross-section}
$$K = \{x_1 = c_1, x_2 = c_2, \ldots, x_r = c_r\}, $$
where $c_i, i=1,\ldots, r$ are some constants and solving the corresponding {\em normalization equations}
\begin{equation}
\omega_1({\bf g},{\bf x}) = c_1, \quad \omega_2({\bf g},{\bf x}) = c_2, \quad \ldots, \quad \omega_r({\bf g},{\bf x}) = c_r,
\label{ne}
\end{equation}
for the group $G$ locally parametrized by ${\bf g} = (g_1,\ldots, g_r)$ in terms of the local coordinates $(x_1,\ldots, x_n)$. 
Substituting the resulting expressions for $g_1,\ldots, g_r$ in terms of the local coordinates $(x_1,\ldots x_n)$ into
the remaining $n-r$ formulas for the transformation rules $\omega({\bf g},{\bf x}) = {\bf g}\cdot {\bf x}$ yields a complete set
of fundamental invariants for the action of $G$ on $M$. 
\begin{theo}
\label{TFI}
If ${\bf g} = \rho({\bf x})$ is the moving frame solution to the normalization equations (\ref{ne}), then the 
functions
\begin{equation}
\label{fi}
{\cal I}_1({\bf x}) = \omega_{r+1}(\rho({\bf x}),{\bf x}),\ldots, {\cal I}_{n-r}({\bf x}) = \omega_{n}(\rho({\bf x}),{\bf x})
\end{equation}
form a complete system of functionally independent fundamental $G$-invariants.
\end{theo} 

Let us now illustrate the procedure and demonstrate how the method of moving frames can be effectively applied to the problem of the
determination of the fundamental invariants of the isometry group in the invariant theory of Killing tensors. 

\begin{exa}{\rm Consider the extended vector space ${\cal K}^2(\mathbb{R}^2)\times \mathbb{R}^2$. The corresponding extended parameter
space $\Sigma \times \mathbb{R}^2$ is determined by the parameters $\beta_1,\ldots, \beta_6, x,y$, where $\beta_i,$ $i=1,\ldots, 6$ are
as in (\ref{gKt}) and $x,y$ are the standard Cartesian coordinates. The isometry group $I(\mathbb{R}^2)$ acting on ${\cal K}^2(\mathbb{R}^2)\times
\mathbb{R}^2$ induces the corresponding transformations on the extended  parameter space $\Sigma\times\mathbb{R}^2$ (\ref{TLC}), which in 
this case take the following form. 

\begin{equation}
\begin{array}{rcl}
\tilde{\beta_1} &=& \beta_1\cos^2\theta - 2\beta_3\cos\theta\sin\theta + \beta_2\sin^2\theta - 2b\beta_4\cos\theta - 2b\beta_5\sin\theta
\\[0.3cm] 
                  & & + \beta_6b^2, \\ [0.3cm]
\tilde{\beta_2} & =& \beta_1\sin^2\theta - 2\beta_3\cos\theta\sin\theta +
\beta_2\cos^2\theta - 2a\beta_5\cos\theta + 2a\beta_4\sin\theta \\ [0.3cm]
 & & + \beta_6a^2,\\[0.3cm] 
\tilde{\beta_3} & = & (\beta_1-\beta_2)\sin\theta\cos\theta + \beta_3(\cos^2\theta - \sin^2\theta) 
+ (a\beta_4 + b\beta_5)\cos\theta \\ [0.3cm] 
 & &  + (a\beta_5 - b\beta_4)\sin\theta - \beta_6ab, \\[0.3cm]
 \tilde{\beta_4} & = & \beta_4\cos\theta + \beta_5\sin\theta - \beta_6b, \\[0.3cm] 
 \tilde{\beta_5} & = & \beta_5\cos\theta - \beta_4\sin\theta - \beta_6a,\\[0.3cm]
 \tilde{\beta_6} &=& \beta_6, \\[0.3cm]
 \tilde{x} &=& x\cos\theta -y\cos\theta +a,\\[0.3cm]
 \tilde{y} & =&  x\sin\theta + y\cos\theta +b.		  
\end{array}
\label{TLC1}
\end{equation} 

Next, we construct a moving frame by using the 
cross-section (for example) 
\begin{equation}
K = \{\beta_3 = \beta_4 = \beta_5 = 0  \},
\label{cs} 
\end{equation}
which yields the corresponding normalization equations
\begin{equation}
\begin{array}{rcl}
0 & = & (\beta_1-\beta_2)\sin\theta\cos\theta + \beta_3(\cos^2\theta - \sin^2\theta) 
+ (a\beta_4 + b\beta_5)\cos\theta \\ [0.3cm] 
 & &  + (a\beta_5 - b\beta_4)\sin\theta - \beta_6ab, \\[0.3cm]
0 & = & \beta_4\cos\theta + \beta_5\sin\theta - \beta_6b, \\[0.3cm] 
0 & = & \beta_5\cos\theta - \beta_4\sin\theta - \beta_6a.
\end{array}
\label{ne1}
\end{equation}
Solving (\ref{ne1}) for the parameters $a,b$ and $\theta$, we obtain the moving frame map $\rho:\,\Sigma \times \mathbb{R}^2 \rightarrow I(\mathbb{R}^2)$ 
determined by the following formulas:
\begin{equation}
\begin{array}{rcl}
a & = & \displaystyle \frac{\beta_5\cos\theta - \beta_4\sin\theta}{\beta_6}, \\ [0.3cm]
b & = & \displaystyle \frac{\beta_4\cos\theta + \beta_5\sin\theta}{\beta_6}, \\ [0.3cm]
\theta &=& \displaystyle \frac{1}{2}\arctan \frac{2(\beta_3\beta_6 +\beta_4\beta_5)}{\beta_6(\beta_1-\beta_2) - \beta^2_4+\beta^2_5}.
\end{array}
\label{mf}
\end{equation}
}
\label{EC1}
\end{exa}

It was observed in \cite{DHMS} that the method of moving frames could be used to solve the problem of the determination of fundamental invariants
of vector spaces of Killing tensors under the action of the isometry group. Indeed, having derived the moving frame map (\ref{mf}) and the transformation
laws (\ref{TLC1}), we can now make use of the result of Theorem \ref{TFI} and determine a set of fundamental $I(\mathbb{R}^2)$-covariants of
 ${\cal K}^2(\mathbb{R}^2)$. Substituting (\ref{mf}) into (\ref{TLC1}), by Theorem \ref{TFI}, we arrive
at the following result. 
\begin{theo} \label{TC1} Consider the vector space  ${\cal K}^2(\mathbb{R}^2)$. Any algebraic $I(\mathbb{R}^2)$-covariant
$C$ defined over the subspace of $\Sigma\times \mathbb{R}^2$ where the isometry group $I(\mathbb{R}^2)$ acts freely and regularly 
with three-dimensional orbits can be locally uniquely expressed as an analytic function $$C= F({\cal I}_1, {\cal I}_2, {\cal I}_3, {\cal C}_1, {\cal C}_2),$$
where the fundamental $I(\mathbb{R}^2)$-covariants ${\cal I}_i$, ${\cal C}_j$, $i=1,2,3$, $j =1,2$ are given by
\begin{equation}
\begin{array}{rcl}
\label{FC}
{\cal I}_1 &=& [\beta_6(\beta_1-\beta_2) + \beta_5^2 - \beta_4^2]^2 +
4(\beta_3\beta_6 +\beta_4\beta_5)^2, \\[0.3cm]
{\cal I}_2 & = & \beta_6(\beta_1+\beta_2) - \beta_4^2 - \beta_5^2, \\[0.3cm]
{\cal I}_3 & = & \beta_6, \\[0.3cm]
{\cal C}_1 & = &  (\beta_6x + \beta_5)^2 + (\beta_6y + \beta_4)^2 \\[0.3cm]
{\cal C}_2 & = &  [(\beta_6x + \beta_5)^2 - (\beta_6y + \beta_4)^2](\beta_5^2-\beta_4^2 + \beta_6(\beta_1-\beta_2)) \\[0.3cm]
& & + 4(\beta_6x + \beta_5)(\beta_6y + \beta_4)(\beta_6\beta_3 + \beta_4\beta_5),
\end{array} 
\end{equation}
where $\Sigma$ is the parameter space of ${\cal K}^2(\mathbb{R}^2).$
\end{theo} 
We immediately observe that the functions ${\cal I}_1, {\cal I}_2, {\cal I}_3$ constitute  in fact a set of fundamental $I(\mathbb{R}^2)$-invariants
of the vector space ${\cal K}^2(\mathbb{R}^2)$, while the functions ${\cal C}_1$ and ${\cal C}_2$ are ``truly'' fundamental $I(\mathbb{R}^2)$-covariants
of the vector space ${\cal K}^2(\mathbb{R}^2)$. We also observe that the fundamental covariant ${\cal C}_1$ can be expressed as 
$$ {\cal C}_1 = {\cal I}_3\mbox{tr}\hat{\bf K} - {\cal I}_2,$$
where the  $(1,1)$-tensor $\hat{\bf K}$ is given by $\hat{\bf K} = {\bf K}{\bf g}^{-1}$. This observation immediately suggests that
$\mbox{tr}\hat{\bf K}$ is a fundamental $I(\mathbb{R}^2)$-covariant of ${\cal K}^2(\mathbb{R}^2)$. We note, however, that the function $\mbox{det}\hat{\bf K}$
is not a  fundamental $I(\mathbb{R}^2)$-covariant of ${\cal K}^2(\mathbb{R}^2)$. 

Consider a similar example.
\begin{exa} \label{EC2} 
{\rm Let ${\cal K}^2(\mathbb{R}_1^2)\times \mathbb{R}_1^2$ be the
extended vector space of ${\cal K}^2(\mathbb{R}_1^2)$. 
The action of the isometry group $I(\mathbb{R}_1^2)$ in the Minkowski plane
$\mathbb{R}_1^2$ is given by (\ref{RM}), while the corresponding action in the parameter space
$\Sigma$ of ${\cal K}^2(\mathbb{R}_1^2)$ is given by (\ref{transf}). The trasformation laws (\ref{transf}) combined with 
the trasformations (\ref{RM}) yield an analogue of (\ref{TLC1}). Next, we proceed as in Example \ref{EC1}. The resulting
moving frame map $\rho:\,\Sigma \times \mathbb{R}^2_1 \rightarrow I(\mathbb{R}^2_1)$ is given by
\begin{equation}
\begin{array}{rcl}
a & = & \displaystyle \frac{\alpha_4\sinh\phi + \alpha_5\cosh\phi}{\alpha_6}, \\ [0.3cm]
b & = & \displaystyle \frac{\alpha_4\cosh \phi + \alpha_5\sinh\phi}{a_6}, \\ [0.3cm]
\phi &=& \displaystyle   \frac{1}{2}\arctanh\frac{2(\alpha_3\alpha_6 - \alpha_4\alpha_5)}{\alpha_4^2 + \alpha_5^2 - \alpha_6(\alpha_1+\alpha_2)}.
\end{array}
\label{mf1}
\end{equation}
}
\end{exa}

Now we can continue as in the previous example to determine a set of fundamental $I(\mathbb{R}_1^2)$-covariants of the vector space ${\cal K}^2(\mathbb{R}_1^2)$.
\begin{theo} \label{TC2}
Consider the vector space ${\cal K}^2(\mathbb{R}^2_1)$. Any algebraic $I(\mathbb{R}^2_1)$-covariant
$C$ defined over the subspace of $\Sigma\times \mathbb{R}^2_1$ where the isometry group $I(\mathbb{R}^2_1)$ acts freely and regularly 
with three-dimensional orbits can be locally uniquely expressed as an analytic function $$C= F({\cal I}_1, {\cal I}_2, {\cal I}_3, {\cal C}_1, {\cal C}_2),$$
where the fundamental $I(\mathbb{R}^2_1)$-covariants ${\cal I}_i$, ${\cal C}_j$, $i=1,2,3$, $j =1,2$ are given by
\begin{equation}
\begin{array}{rcl}
\label{FC1}

{\cal I}_1 & = & [\alpha_4^2+\alpha_5^2-\alpha_6(\alpha_1+\alpha_2)]^2 -4(\alpha_3\alpha_6 -\alpha_4\alpha_5)^2, \\ [0.3cm]

{\cal I}_2 & = & (\alpha_1-\alpha_2)\alpha_6 - \alpha_4^2+\alpha_5^2  \\ [0.3cm]

{\cal I}_3 & = & \alpha_6\\ [0.3cm]

{\cal C}_1 & = &  (\alpha_6t + \alpha_5)^2 - (\alpha_6x + \alpha_4)^2 \\[0.3cm]
{\cal C}_2 & = &  [(\alpha_6t + \alpha_5)^2 + (\alpha_6x + \alpha_4)^2](\alpha_4^2 + \alpha_5^2 - \alpha_6(\alpha_1+\alpha_2)) \\[0.3cm]
& & + 4(\alpha_6t + \alpha_5)(\alpha_6x + \alpha_4)(\alpha_3\alpha_6 - \alpha_4\alpha_5),
\end{array} 
\end{equation}
where $\Sigma$ is the parameter space of ${\cal K}^2(\mathbb{R}^2_1).$
\end{theo} 
The conclusion is similar to that following Theorem \ref{TC1}. Thus, we observe again that  the functions ${\cal I}_1, {\cal I}_2, {\cal I}_3$ constitute  in
fact a set of fundamental $I(\mathbb{R}^2_1)$-invariants
of the vector space ${\cal K}^2(\mathbb{R}^2_1)$, while the functions ${\cal C}_1$ and ${\cal C}_2$ are ``truly'' fundamental $I(\mathbb{R}^2_1)$-covariants
of the vector space ${\cal K}^2(\mathbb{R}^2_1)$. 

\section{Equivalence classes of vector spaces $\bf {\cal K}^2({\mathbb R}^2)$ and $\bf {\cal K}^2({\mathbb R}_1^2)$}

In this section we use the results obtained in the previous section to solve the problems of equivalence for the vector subspaces of {\em
non-trivial} Killing tensors of
$\bf {\cal K}^2({\mathbb R}^2)$ and $\bf {\cal K}^2({\mathbb R}_1^2)$. As is well-known \cite{Be} the elements of these subspaces
generate {\em orthogonal coordinate webs} in $\mathbb{R}^2$ and $\mathbb{R}_1^2$ respectively, provided the Killing tensors
in question have distinct (and real) eigenvalues. The problem of equivalence in this case is the 
problem of classification of orthogonal coordinate webs. On the other hand, from the invariant theory point of view
the problem of equivalence and the related canonical form
problem are intimately related to the problem of the determination of funamental invariants (covariants, joint invariants). 

\subsection{The vector space ${\cal K}^2({\mathbb R}^2)$}

Let ${\cal K}_{nt}^2(\mathbb{R}^2) \subset {\cal K}^2({\mathbb R}^2)$ be the vector subspace of non-trivial Killing two tensors
defined in the Euclidean plane $\mathbb{R}^2$. ``Non-trivial'' in this context means that
none of the elements of ${\cal K}_{nt}^2(\mathbb{R}^2)$ is a multiple of the metric of $\mathbb{R}^2$. 
Clearly $\dim\,{\cal K}_{nt}^2(\mathbb{R}^2) =5.$   It has been established in \cite{MST4, MST5, MST1} that the functions ${\cal I}_1$ and ${\cal I}_3$
 given by (\ref{FC})
are the fundamental $I(\mathbb{R}^2)$-invariants of ${\cal K}_{nt}^2(\mathbb{R}^2)$. Moreover, 
they can be used to solve the problem of classification of orthogonal coordinate webs in the Euclidean plane. The fundamental $I(\mathbb{R}^2)$-invariants
divide the vector subspace ${\cal K}_{nt}^2(\mathbb{R}^2)$ into four equivalence classes. The elements within each eaquivalence class
generate a particular orthogonal web (see \cite{MST4} for more details). These
results are summarized in Table \ref{T1}. Clearly, any (analytic) $I(\mathbb{R}^2)$-covariant of the vector
subspace ${\cal K}_{nt}^2(\mathbb{R}^2)$ takes the following general form.
$$C= F({\cal I}_1, {\cal I}_3, {\cal C}_1, {\cal C}_2),$$
where the functions ${\cal I}_1,$ ${\cal I}_3$, ${\cal C}_1$ and ${\cal C}_2$ are given by (\ref{FC}). 

 The same classification can be done by means of the fundamental $I(\mathbb{R}^2)$-covariants ${\cal C}_1$
and ${\cal C}_2$ given by (\ref{FC}). The results are summarized in Table
\ref{T2}.

 Recall that in most of the problems studied so far within ITKT the associated {\em  canonical form problem }
 has been solved for vector spaces of Killing tensors of valence two via 
 transforming the corresponding Killing tensors in  orthogonal
 coordinates  back to the original (pseudo-) Cartesian coordinates by using
  the standard transformations from the orthogonal coordinates to  (pseudo-) Cartesian
 coordinates (see, for example \cite{HMS, MST3, MST4, MST1}. In the problems involving 
 Killing tensors of valence two (with distinct eigenvalues and integrable eigenvectors) the equivalence classes (ECs) of the corresponding
 vector spaces are associated with  the corresponding orthogonal coordinate webs 
 and so such an approach seems to be natural.
 
 However, one may wish to solve the canonical form problem for vector spaces of
 Killing tensors of valences higher than two, in which case  a connection with the theory of orthogonal coordinate webs
  is not evident. In such a case, another, more general
 approach can be adapted from  CIT \cite{Olv} to the 
 study of Killing tensors.   Indeed, recall first the following definitions and results \cite{Olv}. 
 
 \begin{defi} Two  submanifolds $N, P \subset X$ are said to intersect
 {\em transversally}  at a common point $x_0\in N\bigcap  P$ if they have no
 nonzero tangent vectors in common: $TN|_{x_0}\bigcap TP|_{x_0} = \{0\}$. 
\end{defi} 

\begin{defi}
Let $G$ be a Lie transformation group  that acts regularly on an $m$-dimensional manifold $X$
with $s$-dimensional orbits. A (local) {\em cross-section} is an $(m-s)$-dimensional 
submanifold $K \subset X$ such that $K$ intersects each orbit transversally and at 
most once. 

\end{defi}

\begin{prop} If a Lie group $G$ acts regularly on a manifold $X$, then one can
construct a local cross-section $K$ passing through any point $x\in X$. 

\end{prop}
One can define a {\em coordinate cross-section} $K$, in which case the first $s$
coordinates themselves define a coordinate cross-section \cite{Olv} 
\begin{equation}
\label{K}
K = \{x_1 = c_1, \ldots, x_s = c_s\}
\end{equation} 
iff
\begin{equation}
 \frac{\partial (\Delta_1, \ldots, \Delta_{m-s})}{\partial (x_{s+1},\ldots, x_m)} \not= 0,
 \label{cond}
 \end{equation} 
where $\Delta_1, \ldots \Delta_{m-s}$ are the fundamental invariants of the group action. Then, in view
of the above,  
we can obtain canonical forms of the equivalence classes set by the fundamental invariants as
intersections of the coordinate cross-sections and the level sets (invariant submanifolds) 
defined by the fundamental group invariants. To illustrate
this simple procedure consider the following example. 
\begin{exa} {\rm Consider ${\cal K}_{nt}^2(\mathbb{R}^2) \subset {\cal K}^2(\mathbb{R}^2)$. Without loss of
generality we can assume that  the elements of the vector subspace ${\cal K}_{nt}^2(\mathbb{R}^2)$
enjoy the following general form. 
\begin{equation}
\begin{array}{rcl}
{\bf K}_{nt}^2 & = & \displaystyle (\beta'_1 + 2\beta_4y + \beta_6y^2)\frac{\partial}{\partial x}
\odot  \frac{\partial}{\partial x} \\ [0.3cm]
& & + \displaystyle (\beta_3 - \beta_4x-\beta_5y - \beta_6xy)\frac{\partial}{\partial x}\odot
\frac{\partial}{\partial y} \\ [0.3cm]
& & + \displaystyle (2\beta_5x+\beta_6x^2) \frac{\partial}{\partial y}\odot
\frac{\partial}{\partial y},
\end{array}
\label{Q}
\end{equation}
where $\beta'_1 = \beta_1-\beta_2$ and the parameters $\beta_i,$ $i=1,\ldots, 6$ are as in (\ref{gKt}). 
The four equivalence classes EC1-4 of ${\cal K}_{nt}^2(\mathbb{R}^2)$ have been classified in Table \ref{T1} and 
Table \ref{T2}.  The Killing tensors within each equivalence class share the same geometrical properties, that is they define
the same orthogonal coordinate webs equivalent up to the action of the isometry group $I({\mathbb R}^2)$. 
This fact can be used to select appropriate canonical forms for each of the four equivalence classes. 
Thus, one can consider the Killing tensors in terms of the orthogonal coordinates $(u,v)$ (see \cite{MST1}) and then 
use  the standard coordinate transformations from 
the orthogonal $(u,v)$ coordinates to the  Cartesian coordinates $(x,y)$ in order to determine 
the corresponding canonical forms for EC1-4. Alternatively, one can proceed
by using the coordinate cross-sections. The procedure is outlined below.

\begin{itemize}

\item[EC1.]  In this case the parameter space $\Sigma'$ defined by the five parameters of (\ref{Q}) can be intersected by the coordinate
cross-section
\begin{equation}
\label{K1}
K_1 = \{\beta_3 = \beta_4 = \beta_5 = 0\}
\end{equation} 
Taking into account (\ref{Q})  and the corresponding formulas for ${\cal I}_1$ and ${\cal I}_3$ given by (\ref{FC}), we conclude that all but one ($\beta'_1$) parameters vanish in this case. 
The parameter $\beta'_1$ is arbitrary, without loss of generality we can set $\beta'_1=1$, which leads to the
canonical form 
\begin{equation} \label{C1} 
{\bf K}_I =  \frac{\partial}{\partial x}
\odot  \frac{\partial}{\partial x} 
\end{equation} 
Alternatively, we could have used the coordinate cross-section
\begin{equation}
\label{K2} 
K_2 = \{\beta'_1 = \beta_4 = \beta_5 =  0\}, 
\end{equation} 
which would have led to the canonical form 
\begin{equation} \label{C2} 
{\bf K}'_I =  \frac{\partial}{\partial x}
\odot  \frac{\partial}{\partial y}. 
\end{equation} 
Note the canonical forms (\ref{C1}) and (\ref{C2}) are equivalent
up to a rotation. 

\item[EC2.]  Reason as in EC1 above. Either of the coordinate cross-sections
(\ref{K1}) and (\ref{K2}) leads to the canonical form 
\begin{equation} \label{C3} 
{\bf K}_{II} =  y^2\frac{\partial}{\partial x}
\odot  \frac{\partial}{\partial x}  -xy \frac{\partial}{\partial x}
\odot  \frac{\partial}{\partial y} +  x^2 \frac{\partial}{\partial y}
\odot  \frac{\partial}{\partial y}. 
\end{equation} 

\item[EC3.] First, note that the condition ${\cal I}_1 \not= 0,$ ${\cal I}_3 = 0$ (see Table \ref{T1}) prompts $\beta_4^2 + \beta_5^2
\not=0$. Therefore the coordinate cross-sections that can be used in 
this case are: 
\begin{equation}
\label{K3}
K_3 = \{\beta'_1 = \beta_3 = \beta_4= 0\}
\end{equation} 
and 
\begin{equation}
\label{K4}
K_4 = \{ \beta'_1 = \beta_3 = \beta_5= 0 \}, 
\end{equation}
which lead to the canonical forms 
\begin{equation} \label{C4} 
{\bf K}_{III} =    -y \frac{\partial}{\partial x}
\odot  \frac{\partial}{\partial y} +  2x \frac{\partial}{\partial y}
\odot  \frac{\partial}{\partial y} 
\end{equation}
and
 \begin{equation} \label{C5} 
{\bf K}'_{III} =  2y\frac{\partial}{\partial x}
\odot  \frac{\partial}{\partial x}  -x\frac{\partial}{\partial x}
\odot  \frac{\partial}{\partial y} 
\end{equation}
respectively.
Note the canonical forms (\ref{C4}) and (\ref{C5}) are equivalent up 
to a rotation. 

\item[EC4.] In this case we can use either of the coordinate cross-sections
(\ref{K1}) and (\ref{K2}). Intersecting the common level set defined by ${\cal I}_1 \not= 0,$ ${\cal I}_3 \not= 0$ (see Table \ref{T1}) with
(\ref{K1}) yields the canonical form 
\begin{equation} \label{C6} 
{\bf K}_{IV} =    (\beta'_1+y^2)\frac{\partial}{\partial x}
\odot  \frac{\partial}{\partial x}  -xy \frac{\partial}{\partial x}
\odot  \frac{\partial}{\partial y} + x^2  \frac{\partial}{\partial y}
\odot  \frac{\partial}{\partial y},
\end{equation}
while with (\ref{K2}) - the canonical form 
\begin{equation} \label{C7} 
{\bf K}'_{IV} =    y^2\frac{\partial}{\partial x}
\odot  \frac{\partial}{\partial x} + (\beta_3-xy) \frac{\partial}{\partial x}
\odot  \frac{\partial}{\partial y} + x^2  \frac{\partial}{\partial y}
\odot  \frac{\partial}{\partial y}. 
\end{equation}
Note the canonical forms (\ref{C6}) and (\ref{C7}) are equivalent up to 
a rotation and rescaling. 

\end{itemize} 
}
\end{exa} 

\subsection{The vector space ${\cal K}^2({\mathbb R}^2_1)$}
The problem of classification of the ten orthogonal coordinate webs defined in the Minkowski plane
$\mathbb{R}_1^2$ was initially solved by Kalnins \cite{Kal} in 1975. The approach used in \cite{Kal} is based on
the property that the Killing tensors defined in pseudo-Riemannian spaces of constant curvature are
the sums of symmetrized tensor products of Killing vectors. In \cite{Kal} different combinations (as symmetric tensor products) of the basic Killing vectors (\ref{TXH})
were analysed modulo the action of the eight-demensional discrete group ${\cal R}$ of permutations of coordinates and  reflections of the
signature of the Minkowski metric ${\bf g} = \diag(1,-1)$ given in terms of the pseudo-Cartesian coordinates $(t,x)$ (see below). 
A different approach was used in Rastelli \cite{Rast}, where the ten orthogonal webs were classified based on the algebraic properties
of the non-trivial Killing tensors of ${\cal K}^2({\mathbb R}^2_1)$. More specifically, the author made use of the points where the eigenvalues of such
Killing tensors coincide (singular points). Finally, McLenaghan {\em et al} \cite{MST3,MST6}
employed a set of the fundamental $I(\mathbb{R}_1^2)$-invariants of the vector subspace of non-trivial Killing tensors of 
${\cal K}^2({\mathbb R}^2_1)$ to classify the ten orthogonal webs defined in $\mathbb{R}_1^2$. The problem appeared to 
be incommensurably more challenging than the problem of classification of the orthogonal coordinate webs in $\mathbb{R}^2$ \cite{MST4,MST1}. 
The reason is simple: In both cases one has two fundamental invariants at one's disposal, while the number of orthogonal coordinate webs
is four (Euclidean plane) and ten (Minkowski plane). In the latter case the problem was solved \cite{MST3,MST6} by introducing the concept of
a {\em conformal $I(\mathbb{R}_1^2)$-invariant}, which was used to generate additional {\em discrete} $I(\mathbb{R}_1^2)$-invariants. To solve the problem, 
the authors had to investigate the effect of the eight dimensional discrete group ${\cal R}$ on the discrete $I(\mathbb{R}_1^2)$-invariants. 
Unordered pairs (as the objects preserved by the discrete group) of discrete invariants along with one of the fundamental invariants were used to solve the problem. In what follows, we propose a simpler
solution based on the fundamental $I(\mathbb{R}_1^2)$-covariants obtained in the previous section. 

Let ${\cal K}_{nt}^2(\mathbb{R}^2_1) \subset {\cal K}^2({\mathbb R}^2_1)$ be the vector subspace of non-trivial Killing two tensors
defined in the Minkowski plane $\mathbb{R}^2_1$. Here  "non-trivial" has the same meaning as above. Again $\dim\, {\cal K}_{nt}^2(\mathbb{R}^2_1) =5$.
Without loss of generality we can assume that in terms of the pseudo-Cartesian coordinates $(t,x)$
 the general form of the elements of ${\cal K}_{nt}^2(\mathbb{R}^2_1)$ is given by
\begin{equation}
\begin{array}{rcl}
{\bf K} & = & \displaystyle (\alpha'_1 + 2\alpha_4x + \alpha_6x^2)\frac{\partial}{\partial t}
\odot  \frac{\partial}{\partial t} \\ [0.3cm]
& & + \displaystyle (\alpha_3 + \alpha_4t +\alpha_5x  +  \alpha_6tx)\frac{\partial}{\partial t}\odot
\frac{\partial}{\partial x} \\ [0.3cm]
& & + \displaystyle (2\alpha_5t+\alpha_6t^2) \frac{\partial}{\partial x}\odot
\frac{\partial}{\partial x},
\end{array}
\label{ngKT}
\end{equation}
where $\alpha'_1 = \alpha_1+\alpha_2$ and the parameters $\alpha_i,$ $i=1,\ldots, 6$ are as in (\ref{gKT}). Note
that in this case the parameter space $\Sigma'$ is determined by the five parameters $\alpha'_1,$ $\alpha_3,$ $\alpha_4,$ $\alpha_5$ and
$\alpha_6$. Our next observation is that by Theorem \ref{TC2} any $I(\mathbb{R}_1^2)$-covariant of ${\cal K}_{nt}^2(\mathbb{R}^2_1)$
enjoys the form
$$C= F({\cal I}_1, {\cal I}_3, {\cal C}_1, {\cal C}_2),$$
where the functions ${\cal I}_1,$ ${\cal I}_3$, ${\cal C}_1$ and ${\cal C}_2$ are given by (\ref{FC1}). As in the case of ${\cal K}_{nt}^2(\mathbb{R}^2)$
we can use ${\cal I}_1,$ ${\cal I}_3$, ${\cal C}_1$ and ${\cal C}_2$ to classify the ten orthogonal webs. However, in view of the number of
cases we have to use these functions concurrently. Before doing so, we  
check the effect of ${\cal R}$ on ${\cal I}_1,$ ${\cal I}_3$, ${\cal C}_1$ and
${\cal C}_2$. Recall \cite{Kal, MST3} that the group (under composition) 
${\cal R} = <R_1, R_2>$ consists of eight discrete transformations generated by
\begin{equation}
\begin{array}{rlll}
R_1: & \tilde{t} = t, & \tilde{x} = -x & \mbox{(spatial reflections),} \\[0.3cm]
R_2:  & \tilde{t} = x , & \tilde{x} = t & \mbox{(permutation)}.
\end{array}
\end{equation}
Note the group ${\cal R}$ (along with the isometry group $I(\mathbb{R}_1^2)$) preserves the
geometry of the ten orthogonal webs defined in the Minkowski plane. Recall next \cite{MST3} that
$R_1$ and $R_2$ induce the following transformations on the parameters $\alpha_i,$ $i=1,\ldots, 6$
of ${\cal K}^2(\mathbb{R}^2_1)$ (see (\ref{gKt})): 
\begin{equation}
\begin{array}{rllllll}
\label{RR}
R_1: & \tilde{\alpha}_1 = \alpha_1, & \tilde{\alpha}_2 = \alpha_2, & \tilde{\alpha}_3 = -\alpha_3,& 
\tilde{\alpha}_4=-\alpha_4, & \tilde{\alpha}_5 = \alpha_5, & \tilde{\alpha}_6 = \alpha_6, \\ [0.3cm]
R_2: & \tilde{\alpha}_1 = \alpha_2, & \tilde{\alpha}_2 = \alpha_1, & \tilde{\alpha}_3 = \alpha_3,& 
\tilde{\alpha}_4=\alpha_5, & \tilde{\alpha}_5 = \alpha_4, & \tilde{\alpha}_6 = \alpha_6.
\end{array}
\end{equation}
It follows immediately that the fundamental $I(\mathbb{R}_1^2)$-covariants ${\cal I}_1,$ ${\cal I}_3$, ${\cal C}_1$ and ${\cal C}_2$
remain unchanged under the transformations (\ref{RR}) induced by the group ${\cal R}$. We conclude therefore that we can use
them in the classification of the ten orthogonal webs. Recall that the vector subspace ${\cal K}_{nt}^2(\mathbb{R}^2_1)$ can be divided
into ten equivalence classes EC1-10 whithin each of which the corresponding elements generate the {\em same orthogonal coordinate web} (for more
details see \cite{Kal, MST3}). We consider next the ten {\em canonical elements} determined in \cite{MST3} representing each class
EC1-10 by transforming them to contravariant form and making them compatible with the general form (\ref{ngKT}) by adding multiples
of the metric when necessary. The latter operation does not affect the geometry of the coordinate webs generated by the canonical elements.
We arrive at the following list. 
\begin{equation}
\label{EC11}
\begin{array}{rlcl}
\mbox{EC1} & {\bf K}_1 & = & \displaystyle \frac{\partial}{\partial t}\odot \frac{\partial}{\partial t}, 
\end{array}
\end{equation}
\begin{equation}
\label{EC21}
\begin{array}{rlcl}
\mbox{EC2} & {\bf K}_{2} & = & \displaystyle x^2\frac{\partial}{\partial t}\odot\frac{\partial}{\partial t} + 
tx \frac{\partial}{\partial t}\odot\frac{\partial}{\partial x} + t^2\frac{\partial}{\partial x}\odot\frac{\partial}{\partial x},
\end{array}
\end{equation}
\begin{equation}
\label{EC3}
\begin{array}{rlcl}
\mbox{EC3} & {\bf K}_{3} & = & \displaystyle \Big(\frac{1}{2} - x\Big) \frac{\partial}{\partial t}\odot\frac{\partial}{\partial t} + 
\Big(\frac{1}{4}- \frac{1}{2}t + \frac{1}{2}x\Big)\frac{\partial}{\partial t}\odot\frac{\partial}{\partial x } \\[0.3cm]
& & &  \displaystyle + t\frac{\partial}{\partial x}\odot\frac{\partial}{\partial x}, 
\end{array}
\end{equation}
\begin{equation}
\label{EC4}
\begin{array}{rlcl}
\mbox{EC4} & {\bf K}_{4} & = & \displaystyle 
x \frac{\partial}{\partial t}\odot\frac{\partial}{\partial x} + 2t\frac{\partial}{\partial x}\odot\frac{\partial}{\partial x}, 
\end{array}
\end{equation}
\begin{equation}
\label{EC5}
\begin{array}{rlcl}
\mbox{EC5}&  {\bf K}_{5}& = & \displaystyle \Big(2k^2- \frac{1}{4}x^2\Big) \frac{\partial}{\partial t}\odot\frac{\partial}{\partial t}  
-\frac{1}{4}tx \frac{\partial}{\partial t}\odot\frac{\partial}{\partial x } \\[0.3cm]
 & & &  \displaystyle - \frac{1}{4}t^2\frac{\partial}{\partial x}\odot\frac{\partial}{\partial x}, 
\end{array}
\end{equation}
\begin{equation}
\label{EC6}
\begin{array}{rlcl}
 \mbox{EC6} & {\bf K}_{6} & = & \displaystyle \Big(\frac{1}{4}+ \frac{1}{4} x^2 \Big)\frac{\partial}{\partial t}\odot\frac{\partial}{\partial t} + 
\Big(\frac{1}{4} +\frac{1}{4}tx \Big)\frac{\partial}{\partial t}\odot\frac{\partial}{\partial x } \\[0.3cm]
& & &  \displaystyle + \frac{1}{4} t^2\frac{\partial}{\partial x}\odot\frac{\partial}{\partial x}, 
\end{array}
\end{equation}
\begin{equation}
\label{EC7}
\begin{array}{rlcl}
 \mbox{EC7} & {\bf K}_{7} & = & \displaystyle \Big(-\frac{1}{2}+\frac{1}{4}x^2\Big)  \frac{\partial}{\partial t}\odot\frac{\partial}{\partial t} + 
\Big(- \frac{1}{4} +\frac{1}{4}tx \Big)\frac{\partial}{\partial t}\odot\frac{\partial}{\partial x } \\[0.3cm]
& & &  \displaystyle + \frac{1}{4} t^2\frac{\partial}{\partial x}\odot\frac{\partial}{\partial x}, 
\end{array}
\end{equation}
\begin{equation}
\label{EC8}
\begin{array}{rlcl} 
 \mbox{EC8} & {\bf K}_{8} & = & \displaystyle \frac{1}{4} x^2 \frac{\partial}{\partial t}\odot\frac{\partial}{\partial t} + 
\Big(-k^2 +\frac{1}{4}tx \Big)\frac{\partial}{\partial t}\odot\frac{\partial}{\partial x } \\[0.3cm]
& & &  \displaystyle + \frac{1}{4} t^2\frac{\partial}{\partial x}\odot\frac{\partial}{\partial x}, 
\end{array}
\end{equation}
\begin{equation}
\label{EC9}
\begin{array}{rlcl}
\mbox{EC9}&  {\bf K}_{9}& = & \displaystyle \Big(2k^2 + \frac{1}{4}x^2\Big) \frac{\partial}{\partial t}\odot\frac{\partial}{\partial t}  
+\frac{1}{4}tx \frac{\partial}{\partial t}\odot\frac{\partial}{\partial x } \\[0.3cm]
 & & &  \displaystyle + \frac{1}{4}t^2\frac{\partial}{\partial x}\odot\frac{\partial}{\partial x}, 
\end{array}
\end{equation}
\begin{equation}
\label{EC10}
\begin{array}{rlcl}
\mbox{EC10}&  {\bf K}_{10}& = & \displaystyle \Big(-2k^2 + \frac{1}{4}x^2\Big) \frac{\partial}{\partial t}\odot\frac{\partial}{\partial t}  
+\frac{1}{4}tx \frac{\partial}{\partial t}\odot\frac{\partial}{\partial x } \\[0.3cm]
 & & &  \displaystyle + \frac{1}{4}t^2\frac{\partial}{\partial x}\odot\frac{\partial}{\partial x}, 
\end{array}
\end{equation}
where the parameter $k$ is a $I(\mathbb{R}_1^2)$-invariant of ${\cal K}_{nt}^2(\mathbb{R}_1^2)$. In view of Theorem \ref{TFM} (see
also Theorem \ref{TC2}), it can 
be represented via the fundamental $I(\mathbb{R}_1^2)$-invariants. Indeed, the corresponding formulas were found in \cite{MST3}:
\begin{equation}
\label{k}
\begin{array}{rll}
\mbox{EC5, EC9, EC10}: & k^2 = \displaystyle \frac{\sqrt{{\cal I}_1}}{{\cal I}_3}, & ({\cal I}_1 > 0), \\[0.3cm]
\mbox{EC8}: & k^2 = \displaystyle \frac{\sqrt{-{\cal I}_1}}{{\cal I}_3}, & ({\cal I}_1 < 0).
\end{array}
\end{equation}
Note the canonical forms (\ref{EC11})-(\ref{EC10}) are compatible with the general form given by (\ref{ngKT}). 
Following the procedure devised in \cite{MST3}, we  use the canonical forms (\ref{EC11}-\ref{EC10}) to evaluate the corresponding values of
the fundamental $I(\mathbb{R}^2_1)$-covariants ${\cal I}_1,$ ${\cal I}_3$, ${\cal C}_1,$ ${\cal C}_2$ and employ the results
to distinguish the elements belonging to different equivalence classes EC1-10. The elements of ${\cal K}_{nt}^2(\mathbb{R}^2_1)$
must have the same values of ${\cal I}_1,$ ${\cal I}_3$, ${\cal C}_1$ and ${\cal C}_2$. We note however that these
functions do not distinguish EC1 from EC3 and EC6 from EC8. Therefore we have to derive some auxiliary $I(\mathbb{R}_1^2)$-invariants
to  complete the classification scheme. Indeed, consider the vector space ${\cal K}^2(\mathbb{R}_1^2)$ under the action of
the isometry group $I(\mathbb{R}_1^2)$. Since ${\cal I}_3$ is a fundamental $I(\mathbb{R}_1^2)$-invariant, we can 
consider the level set
\begin{equation}
\label{LS}
 {\cal S}_{{\cal I}_3} = \{(\alpha_1,\ldots, \alpha_5)\in\Sigma\, | \quad {\cal I}_3 = 0\}.
 \end{equation}
Note ${\cal S}_{{\cal I}_3}$ is an  $I(\mathbb{R}_1^2)$-invariant submanifold in $\Sigma$ defined by the parameters $\alpha_i,$
$i = 1, \ldots, 5$. Next we prove  the following result by
using the techniques exhibited in  Section 2. 
\begin{lem} Any algebraic $I(\mathbb{R}^2_1)$-invariant ${ I}$ of the $I(\mathbb{R}_1^2)$-invariant submanifold ${\cal S}_{{\cal I}_3}$ defined 
by (\ref{LS})
can be (locally) uniquely expressed as an analytic function 
$${I}= F({\cal I}'_1, {\cal I}'_2)$$
where the fundamental invariants ${\cal I}'_i,$ $i=1,2$ are given by
\begin{equation}
\label{AI}
\begin{array}{rcl}
{\cal I}'_1 & = & \alpha_4^2 - \alpha_5^2,\\ [0.3cm]
{\cal I}'_2 & = & 2\alpha_3\alpha_4\alpha_5 - \alpha_2\alpha_4^2 - \alpha_1\alpha_5^2,
\end{array}
\end{equation}
provided the group acts in ${\cal S}_{{\cal I}_3}$ with three-dimensional orbits.
\label{AL}
\end{lem} 
We note that the fundamental $I(\mathbb{R}^2_1)$-invariants ${\cal I}'_1$ and ${\cal I}'_2$ still cannot be used in the 
problem of classification of the elements of ${\cal K}_{nt}^2(\mathbb{R}_1^2)$. In particular, ${\cal I}'_2$ appears to be a function
of $\alpha_1,$ $\alpha_2,$ $ \alpha_3,$ $\alpha_4$ and $\alpha_5$ (not $\alpha'_1,$ $\alpha_3,$  $ \alpha_4,$  $ \alpha_5 $). However, under the additional
{\em invariant} condition
\begin{equation}
\label{cond2}
{\cal I}'_1 = \alpha_4^2 - \alpha_5^2 = 0
\end{equation}
it assumes the following form: 
\begin{equation}
{\cal I}'_2 = 2\alpha_3\alpha_4\alpha_5 - \alpha'_1\alpha_4^2,
\label{cond1}
\end{equation}
where $\alpha'_1 = \alpha_1 + \alpha_2.$ We immediately recognize the $I(\mathbb{R}^2_1)$-invariant (\ref{cond1}) to
be an $I(\mathbb{R}^2_1)$-invariant of the submanifold in  ${\cal S}_{{\cal I}_3}$  determined by the condition (\ref{cond2}). 
Hence, ${\cal I}'_2$ given by (\ref{cond1}) can be used to distinguish between EC1 and EC3. 

Next, in order  to distinguish between the elements of  EC6 and EC8, introduce the following auxiliary $I(\mathbb{R}^2_1)$-invariant.
\begin{equation}
{\cal I}^* : = k^4 {\cal I}_3 + {\cal I}_1, 
\label{AI1}
\end{equation}
where $k$ is given by (\ref{k}) (the formula for EC8). We note that ${\cal I}^*$ given by (\ref{AI1}) is an $I(\mathbb{R}^2_1)$-invariant.
The values of ${\cal I}_1$ and ${\cal I}_3$ evaluated with respect to
the parameters of the canonical form EC8 given by (\ref{EC8}) are
$${\cal I}_1 = \displaystyle -\frac{k^4}{4}, \quad {\cal I}_3 = \frac{1}{4}.$$
Therefore the $I(\mathbb{R}^2_1)$-invariant ${\cal I}^* = 0,$ whenever the Killing tensor in question belongs to EC8. 
The classification scheme is now complete. We summarize the results in Table \ref{T3}. 

Using the results obtained we can devise a general algorithm of classification the elements of the vector spaces 
${\cal K}^2(\mathbb{R}^2)$ and ${\cal K}^2(\mathbb{R}_1^2)$. It consists of the following two steps. Let $\bf K$
$\in {\cal K}^2(\mathbb{R}^2)$ (${\cal K}^2(\mathbb{R}_1^2)$).  
\begin{itemize}
\item[(i)] If $\bf K$ has arbitrary constants, decompose $\bf K$ as follows:
\begin{equation}\label{EK} {\bf K} = \ell_0{\bf g} + \sum_{i=1}^5\ell_i{\bf K}_i, \end{equation}
where $\ell_i$ $i = 1,\ldots, 5$ are the arbitrary constants. Note $\sum_{i=1}^5\ell_i{\bf K}_i \in {\cal K}_{nt}^2(\mathbb{R}^2)$ 
(${\cal K}_{nt}^2(\mathbb{R}^2_1$)). Clearly, ${\bf K} \in {\cal K}^2_{nt}(\mathbb{R}^2)$ $({\cal K}^2_{nt}(\mathbb{R}_1^2))$ iff $\ell_0 =0$. 
 
\item[(ii)] Each Killing tensor in the representation (\ref{EK}) represents one of the equivalence classes (and thus, - an orthogonal
coordinate web), provided it has  real eigenvalues
 in the case of the vector space being 
${\cal K}^2(\mathbb{R}_1^2)$. We can determine which one by evaluating 
the corresponding $I(\mathbb{R}^2)$ and $I(\mathbb{R}^2_1)$-invariants and covariants and then using the information
provided in Table \ref{T1} or Table \ref{T2} for the Killing tensors defined in the Euclidean plane and Table \ref{T3} - 
the Minkowski plane.
\end{itemize}
The problem of classification is therefore solved. 
\begin{rem} {\rm We note that EC5 and EC10 are characterized by the same values of the fundamental $I(\mathbb{R}_1^2)$-covariants. It agrees
with the geometry of the corresponding orthogonal webs, namely they determine two distinct coordinate systems that 
cover two disjoint areas of the same space (see Miller \cite{Mi} for more details).
}
\end{rem}

\bigskip

\noindent {\bf Acknowledgements} We wish to thank  Irina Kogan and Alexander Zhalij  for 
 bringing to our attention  the references \cite{SHA} and   \cite{Bo} respectively. We also acknowledge the useful remarks
 by the anonymous referee that have helped to improve the presentation of our results. The research was supported in part by 
a National Sciences and Engineering Research Council of Canada Discovery Grant (RGS) and an
 Izaak Walton Killam  Predoctoral 
Scholarship (JY).

\newpage
\listoftables
\newpage

\begin{table}[ht]  
\begin{center}  
\begin{tabular}{|c|c|c|c|} \hline 
{ Equivalence class} & ${\cal I}_1$ & ${\cal I}_3$ & {\ Orthogonal web}   \\ \hline
EC1 & $0$ & $0$ & { Cartesian} \\ 
EC2 & $0$ & $\not=0$ & {Polar} \\ 
EC3 & $\not=0$ & $0$ & { Parabolic} \\ 
EC4 & $\not= 0$ & $\not= 0$  & { Elliptic-hyperbolic} \\ \hline
\end{tabular} 
  \caption{\small Invariant classification of the orthogonal coordinate webs in $\mathbb{R}^2$ by means of $I(\mathbb{R}^2)$-invariants.}   
\label{T1}   
\end{center}   
\end{table} 
\newpage

\begin{table}[ht]  
\begin{center}  
\begin{tabular}{|c|c|c|c|} \hline 
{ Equivalence class} & ${\cal C}_1$ & ${\cal C}_2$ & {\ Orthogonal web}   \\ \hline
EC1 & $ 0 $ & $0$ & { Cartesian} \\ 
EC2 & positive-definite & $0$ & {Polar} \\ 
EC3 & 1 & $1$ & {Parabolic} \\ 
EC4 & positive-definite & indefinite  & { Elliptic-hyperbolic} \\ \hline
\end{tabular} 
  \caption{\small Invariant classification of the orthogonal coordinate webs in $\mathbb{R}^2$ by means of $I(\mathbb{R}^2)$-covariants.}   
\label{T2}   
\end{center}   
\end{table}

\newpage

\begin{table}[ht]  
\begin{center}  
\begin{tabular}{|c|c|c|c|c|c|c|c|} \hline 
\begin{tabular}{c}  
Equivalence \\ class \end{tabular}
 & ${\cal I}_1$ & ${\cal I}_3$ & ${\cal C}_1$ & ${\cal C}_2$ & ${\cal I}'_1$ & ${\cal I}'_2$ & ${\cal I}^*$     \\ \hline
 EC1 & 0 & 0 & 0 & 0 & 0 & 0 &  \\ 
 EC2 & 0 & $\not= 0$ & indefinite & 0 &  &  &  \\ 
 EC3 & 0 & 0 & 0 & 0 & 0 & $\not=0$ &  \\
 EC4 & $\not=0$ & 0 & 1 & 1 &  &  &  \\
 EC5 & $\not=0$ & $\not=0$ & indefinite & positive-definite &  &  &  \\
 EC6 & $\not=0$ & $\not=0$ & indefinite & indefinite &  &  & $\not=0$ \\
 EC7 & 0 & $\not=0$ & indefinite & positive-definite &  &  &  \\
 EC8 & $\not=0$ & $\not=0$ & indefinite & indefinite &  &  & 0 \\
 EC9 & $\not=0$ & $\not=0$ & indefinite & negative-definite &  &  &  \\
 EC10 & $\not=0$ & $\not=0$ & indefinite & positive-definite &  &  &  \\ \hline
 \end{tabular} 
  \caption{\small Invariant classification of the orthogonal coordinate webs in $\mathbb{R}^2_1$ by means of $I(\mathbb{R}^2_1)$-invariants
  and covariants.}   
\label{T3}   
\end{center}   
\end{table}

\end{document}